\documentclass{article}
\usepackage{geometry}
\usepackage{cite}

\usepackage{amsmath}
\usepackage{algorithmic}
\usepackage{url}

\usepackage{amssymb}
\usepackage{graphicx}
\usepackage{hyperref}
\usepackage{algorithm}
\usepackage{subcaption}
\usepackage{tikz}
\usepackage{tikzscale}
\usepackage{enumerate}
\usepackage[shortlabels]{enumitem}
\usepackage{boldline}
\usepackage{amsthm}

\newtheorem{proposition}{Proposition}

\begin{document}

%
\title{Online Particle Smoothing with Application to Map-matching}
%
%
%

\makeatletter
\def\blfootnote{\xdef\@thefnmark{}\@footnotetext}
\makeatother

\author{Samuel~Duffield
\qquad Sumeetpal S. Singh \\
		University of Cambridge
}

\maketitle

\begin{abstract}
    We introduce a novel method for online smoothing in state-space models that utilises a fixed-lag approximation to overcome the well known issue of path degeneracy.
    Unlike classical fixed-lag techniques that only approximate certain marginals, we introduce an online resampling algorithm, called \textit{particle stitching}, that converts these marginal samples into a full posterior approximation.
    We demonstrate the utility of our method in the context of map-matching, the task of inferring a vehicle's trajectory given a road network and noisy GPS observations. We develop a new state-space model for the difficult task of map-matching on dense, urban road networks.
\end{abstract}


%

%
%
%
%

\section{Introduction}

\blfootnote{
S. Duffield is supported by the UK Engineering and Physical Sciences Research Council (EPSRC) doctoral training award
\par
Contact: sddd2@cam.ac.uk, sss40@cam.ac.uk
}

Widely used in modelling dynamical systems and time series, a state-space model is fully defined by the following distributions for the hidden $\{x_t \}_{t=0}^\infty$ and observed process $\{y_t \}_{t=0}^\infty$
\begin{align*}
	&p(x_0), \\
	&p(x_t|x_{t-1}), &t=1,2,3,\dots& \\
	&p(y_t|x_t), &t=0,1,2,\dots&
\end{align*}
The statistical inference goal of \textit{smoothing} is the task of approximating the joint smoothing distribution (or posterior distribution)
\begin{equation} \label{full_post}
    p(x_{0:T}|y_{0:T}) \propto p(x_0|y_0) \prod_{t=1}^T p(x_t|x_{t-1}) p(y_t|x_t),
\end{equation}
where $x_{0:T} = (x_0, \dots, x_T)$ are latent states to be inferred and $y_{0:T} = (y_0, \dots, y_T)$ are given observations.
\par
For \textit{online smoothing}, we have the additional requirement of being able to quickly and accurately update an approximation of $p(x_{0:T-1}|y_{0:T-1})$ to approximate $p(x_{0:T}|y_{0:T})$ in light of receiving a new observation $y_{T}$.
\par
A popular and powerful approach to inference in generic state-space models is that of particle filtering/smoothing (or sequential Monte Carlo),  see \cite{Chopin2020} for a recent, thorough review. Particle smoothers approximate the joint smoothing distribution with a collection of weighted (or unweighted) particles
\begin{equation*}
	p(x_{0:T} | y_{0:T}) \approx \sum_{i=1}^N w_T^{(i)} \delta_{x_{0:T}}\left(x_{0:T}^{(i)}\right),
\end{equation*}
as opposed to particle filters which are only asked to approximate the filtering marginals $p(x_T|y_{0:T})$.
Existing online particle smoothing approximations either degenerate as the length of the state-space model increases \cite{Olsson2008, Poyiadjis2011} or only target the smoothing marginals $p(x_t| y_{0:T})$ \cite{Kitagawa2001} rather than the joint smoothing distribution. More recently, efficient techniques have been developed to produce online approximations to expectations with respect to $p(x_{t-1}, x_t | y_{0:T})$ \cite{DelMoral2010, Olsson2017}. However, these are integrand specific whereas a particle approximation to the joint smoothing distribution is significantly more useful as mathematical expectations can be calculated over a variety of functions defined over full trajectories, thus providing the user with complete flexibility.

\par


In summary, our motivation is to develop an algorithm that simultaneously satisfies the following requirements

\begin{enumerate}[label={(R\arabic*)}]
    \item \textbf{Joint Smoothing}: The algorithm efficiently approximates $p(x_{0:T}|y_{0:T})$ rather than only marginals. \label{r1}
    \item \textbf{Online}: Our approximation can be quickly updated as it receives new observations.\label{r2}
    \item \textbf{Non-degenerate}: The algorithm avoids the path degeneracy of classical particle filters.\label{r3}
\end{enumerate}

The application of particle smoothing to urban map-matching is particularly motivating. The combination of dense road networks (with frequent intersections) and noisy GPS observations leads to uncertainty over the location and route of vehicles. This is a compelling application of online joint smoothing where it is desirable to represent the route with a diverse collection of particles, where each particles describes a plausible vehicle trajectory.
\par

The contribution of this work is both methodological and applied.
\begin{itemize}
    \item We develop a new online resampling method called \textit{particle stitching} that converts marginal fixed-lag samples, which are non-degenerate, into a full posterior approximation \eqref{full_post}. In doing so we jointly satisfy \ref{r1}-\ref{r3} using particle smoothing.
    
    \item We present a new state-space model for map-matching in dense, urban road networks. We demonstrate the benefits of uncertainty quantification over popular optimisation-based approaches and show the performance of offline particle smoothing can be matched by the online particle smoothers introduced here.
\end{itemize}

%
The rest of the paper is structured as follows. In Section~\ref{sec:rel_work} we discuss related online and offline algorithms for particle smoothing, as summarised in \tablename~\ref{tab:comp_tab} alongside those introduced in this paper. Section~\ref{sec:stitch} describes how to combine blocked samples in a way that is invariant for a fixed-lag joint smoothing distribution and Section~\ref{sec:block_prop} introduces two efficient online methods for generating these blocked samples. Section~\ref{sec:mm} describes the problem of map-matching and Section~\ref{sec:results} demonstrates numerically the benefits of uncertainty quantification as well as the performance of the introduced online particle smoothers and their sensitivity to key parameters. In Section~\ref{sec:conc} we conclude and discuss some potential extensions.

\section{Background and Related Work}
\label{sec:rel_work}

\subsection{Particle Filtering and Path Degeneracy}

A classical particle filter runs a single forward pass, updating particles at every observation. Each update consists of three steps: an optional \textit{resample} step, a \textit{propagation} step and a \textit{weighting} step. These steps are described in Alg. \ref{alg:pf}.
\par

The resampling operation converts a weighted sample into an unweighted sample that likely contains duplicates $\left\{ x^{(i)}, w^{(i)} \right\}_{i=1}^N \rightarrow \left\{ x^{(i)}, \frac1N \right\}_{i=1}^N$. The most common resampling method is multinomial sampling (with replacement)
\begin{align*}
\begin{split}
    \text{Sample} \quad a^{(i)} &\sim \text{Categorical}\left( \left\{w^{(i)}\right\}_{i=1}^N \right),  \hspace{0.3cm} i=1,\dots, N,\\
    \text{Set} \quad x^{(i)} &\leftarrow x^{(a^{(i)})} \quad w^{(i)} \leftarrow \frac1N, \hspace{1.2cm} i=1,\dots, N,
\end{split}
\end{align*}
where $\text{Categorical}( \{w^{(i)}\}_{i=1}^N ) = \sum_{i=1}^N w^{(i)} \delta(i)$ simply draws an index $i$ from the index set $\{1,\dots,N\}$ with probabilities $w^{(i)}$.

\par

\begin{algorithm}
	\caption{Particle Filter}
	\begin{algorithmic}[1]  \label{alg:pf}
		\item Input weighted sample $\{ x_{0:T-1}^{(i)}, w_{T-1}^{(i)} \}_{i=1}^N$ approximating $p(x_{0:T-1}|y_{0:T-1})$ and new observation $y_{T}$.
		\STATE Resample (optional)
		\begin{align*}
			\left\{ x^{(i)}_{0:T-1}, w^{(i)}_{T-1} \right\}_{i=1}^N
			\rightarrow
			\left\{ x^{(i)}_{0:T-1}, \frac1N \right\}_{i=1}^N
		\end{align*} \label{alg:step:pf:res}
		\STATE Propagate by sampling from some proposal distribution \label{alg:step:pf:prop}
		\begin{align}
			x_{T}^{(i)} \sim q\left(  x_{T} \left| x_{T-1}^{(i)}, y_{T} \right. \right), \tag*{$i=1,\dots,N$.}
		\end{align} \label{pf_propagate}
		Append to particle $x_{0:T}^{(i)} = \left(x_{0:T-1}^{(i)}, x_{T}^{(i)} \right)$.
		\STATE Weight and normalise \label{alg:step:pf:weight}
		\begin{align}
			&w^{(i)}_{T} \propto
			\frac{p\left(x_{T}^{(i)} \left| x_{T-1}^{(i)} \right. \right)
				p\left(y_{T} \left| \tilde{x}_{T}^{(i)} \right. \right)}
			{q\left(x_{T}^{(i)}\left| x_{T-1}^{(i)}, y_{T} \right. \right)}
			w^{(i)}_{T-1} \tag*{$i=1,\dots,N$.} 
		\end{align} \label{pf_weight}
		\STATE Output weighted sample $\{ x_{0:T}^{(i)}, w_T^{(i)}\}_{i=1}^N$ approximating $p(x_{0:T}|y_{0:T})$
	\end{algorithmic}
\end{algorithm}

Although most commonly used only to approximate the filtering marginals $p(x_T|y_{0:T})$, the particle filter update as described in Alg. \ref{alg:pf} does provide an asymptotically unbiased approximation to the full smoothing distribution $p(x_{0:T}|y_{0:T})$.
\par
The reason that a classical particle filter is almost never used to approximate the smoothing distribution is due to \textit{path degeneracy}.
Path degeneracy occurs in state-space models with large $T$. For repeated particle filter updates, the early coordinates of particles (e.g. $x_0^{(i)}$) will only be altered in the resampling step. Resampling can only decrease particle diversity (the number of distinct particles) and therefore as $T$ increases, and resampling has occurred with suitable frequency, the particle approximation of $p(x_0|y_{0:T})$ will eventually collapse to just a single particle duplicated $N$ times.
\par

Due to the optional nature of the resampling step, adaptive schemes are often applied, with a popular choice to be to only resample if the \textit{effective sample size},  $\text{ESS}( \{w^{(i)}\}_{i=1}^N ) =1/\sum_{i=1}^N w^{(i) \, 2}$, falls below some threshold. In addition to criteria such as the effective sample size that ensure resampling is only applied when necessary, a variety of techniques have been developed to mitigate the effects of path degeneracy. These include adaptively increasing the number of particles \cite{Elvira2017} as well as more sophisticated resampling schemes \cite{Douc2005, Tiancheng2015, Koppel2021} which aim to only resample particles with negligible weights. These approaches can improve performance over multinomial resampling however still suffer the collapse in the approximation of $p(x_0|y_{0:T})$ for $T$ sufficiently large.
\par
The merits of resampling are well known, in particular resampling is vital in ensuring a diverse particle approximation to the filtering marginals $p(x_T | y_{0:T})$ - often the particle filter's primary task. Indeed, a particle filter without resampling is simply importance sampling on an a space whose dimension increases with every new observation. As a result, the divergence between importance and target distribution increases and the number of particles must increase exponentially \cite{Chatterjee2018} - the addition of resampling means a stable approximation to the filtering marginals is maintained.

\begin{table*}
	\footnotesize
	\centering
	\begin{tabular}{lV{4}c|c|c|c|c|}
		& \begin{tabular}[c]{@{}c@{}}Joint\\ Smoothing\end{tabular} & Online  & \begin{tabular}[c]{@{}c@{}}Path\\ Degeneracy \end{tabular} & \begin{tabular}[c]{@{}c@{}}Fixed-lag\\ Approx.\end{tabular} & Complexity \\
		\hlineB{4}
		
		Particle Filter & \checkmark & \checkmark & \textcolor{red}{\checkmark} & &$N$ \\ \hline
		
		Marginal Fixed-lag \cite{Kitagawa2001} & & \checkmark & \textcolor{red}{For large $L$} & \textcolor{red}{\checkmark} &$N$ \\ \hline
		
		\begin{tabular}[c]{@{}l@{}}Forward Filtering-\\ Backward Smoothing \cite{Doucet2000} \end{tabular} &  &  &  &  & $N^2$ \\ \hline
		
		\begin{tabular}[c]{@{}l@{}}Forward Filtering-\\ Backward Simulation \cite{Godsill2004} \end{tabular} & \checkmark & & & & \begin{tabular}[c]{@{}c@{}}$N^2$\\ $N$ with RS$^\dagger$\end{tabular} \\ \hline
		
		\begin{tabular}[c]{@{}l@{}}Forward Smoothing\\ for Additive Functionals \cite{DelMoral2010}\end{tabular} & & \checkmark & & & $N^2$ \\ \hline
		
		PaRIS \cite{Olsson2017} & & \checkmark & & & $N$ with RS$^\dagger$\\ \hline
		
		Block Sampling \cite{Doucet2006} &  \checkmark &  \checkmark  & \textcolor{red}{\checkmark} & & $LN$ \\ \hline
		
		\begin{tabular}[c]{@{}l@{}} \textcolor{blue} {Online Particle Smoother}  \\ \textcolor{blue} { (Alg.~\ref{alg:joint_fl_pf})}\end{tabular}
		&  \checkmark &  \checkmark  & \textcolor{red}{For large $L$} & \textcolor{red}{\checkmark} & \begin{tabular}[c]{@{}c@{}} $N^2$ \\ $N$ with RS$^\dagger$\end{tabular} \\ \hline

		\begin{tabular}[c]{@{}l@{}} \textcolor{blue} {Online Particle Smoother} \\  \textcolor{blue} {with Backward Simulation}
			\\  \textcolor{blue} {(Alg.~\ref{alg:joint_fl_ffbsi})}\end{tabular} &  \checkmark &  \checkmark  & & \textcolor{red}{\checkmark} & \begin{tabular}{@{}c@{}} $LN^2$ \\ $LN$ with RS$^\dagger$ \end{tabular}\\
		
		\hline
		
	\end{tabular}
	
	\caption{Comparison of particle smoothing algorithms, for number of particles $N$ and fixed-lag parameter $L$.
		\\
		$^\dagger$For these algorithms the rejection sampling technique of \cite{Douc2011} can be applied to obtain linear complexity when a bound for the transition density $p(x_t | x_{t-1})$ is available.
	}
	\label{tab:comp_tab}
\end{table*}

\subsection{Marginal Fixed-Lag}
An alternative approach in mitigating the path degeneracy induced by repeated resampling is to simply stop resampling the early coordinates of particles, as proposed in \cite{Kitagawa2001}. That is, replace the resample step (Alg. \ref{alg:pf}, step \ref{alg:step:pf:res}) with the scheme described in Alg. \ref{alg:partial_res}.
\begin{algorithm}
	\caption{Marginal Fixed-lag Resampling (for $T>L$)}
	\begin{algorithmic}[1]\label{alg:partial_res}
		\STATE Fix $\{ x^{(i)}_{0:T-L-1} \}_{i=1}^N$
		\STATE Resample only recent coordinates
		\begin{align*}
			\left\{ x^{(i)}_{T-L:T}, w^{(i)}_{T} \right\}_{i=1}^N
			\rightarrow
			\left\{ x^{(i)}_{T-L:T}, \frac1N \right\}_{i=1}^N
		\end{align*}
		Stitch arbitrarily $x_{0:T}^{(i)} = (x_{0:T-L-1}^{(i)}, x_{T-L:T}^{(i)})$.
		
	\end{algorithmic}
\end{algorithm}

The justification for freezing $x_{0:T-L-1}^{(i)}$ is that after a certain lag $L$ the smoothing distribution of early coordinates become (approximately) independent of new observations
\begin{equation} \label{fixed_lag_app}
	p(x_{0:t}|y_{0:T}) \approx p(x_{0:t}|y_{0:\min(t+L, T)}),
\end{equation}
this represents a \textit{fixed-lag approximation}.
\par
The major issue with the fixed-lag resampling scheme described in Alg. \ref{alg:partial_res} is that early and recent coordinates of particles are \textit{arbitrarily stitched} together and therefore only provide a particle approximation to the \textit{fixed-lag marginal} smoothing distribution:
\begin{align}
	p^L_{\text{marg}}(x_{0:T}|y_{0:T}) 
	= \prod_{t=0}^{T-L-1} p(x_t|y_{0:t+L}) 
	p(x_{T-L:T}|x_{T-L-1}, y_{T-L:T}).  \label{fixed_lag_m}
\end{align}
As such we lose the ability to take expectations over the joint distribution of early coordinates $x_{0:T-L-1}$.
\par
A more useful particle approximation targets the \textit{fixed-lag joint} smoothing distribution
\begin{align}
	p^L(x_{0:T}|y_{0:T}) = p(x_0|y_{0:L})
	\prod_{t=1}^{T-L-1} p(x_t|x_{t-1}, y_{t:t+L})
	p(x_{T-L:T}|x_{T-L-1}, y_{T-L:T}),
	\label{fixed_lag_j}
\end{align}
which permits expectations over full trajectories $x_{0:T}$.
\par

\subsection{Offline Smoothing}

A popular method for approximating the smoothing distribution $p(x_{0:T}|y_{0:T})$ is that of \textit{forward filtering-backward smoothing} \cite{Doucet2000}. This method stores the output of each particle filter update $\{x_t^{(i)}\}_{i=1}^N \sim p(x_t | y_{0:t})$ and runs a full backward pass that updates the weights based on the backward decomposition
\begin{equation}\label{back_decomp}
    p(x_{t-1}|x_{t}, y_{0:t-1}) = \frac{p(x_{t}| x_{t-1}) p(x_{t-1} | y_{0:t-1})}{p(x_{t} | y_{0:t-1})},  \qquad t=T,\dots,1.
\end{equation}
\par
Similarly, \textit{forward filtering-backward simulation} (FFBSi) \cite{Godsill2004} runs a full backward pass based on the same decomposition, but samples a new ancestor from the particle cloud (according to \eqref{back_decomp}) rather than only updating the weights and thus targets the joint smoothing distribution. Additionally, \cite{Douc2011} showed that rejection sampling can significantly reduce the complexity of this pass in the case that $p(x_{t}|x_{t-1})$ has a tractable upper bound. An implementation of backward simulation is described in, Alg.~\ref{alg:ffbsi}.
\par
Both of these algorithms require a full backward pass in light of every new observation and therefore are not suitable for online smoothing.
\par

\subsection{Online Smoothing}
It was noted in \cite{DelMoral2010} that the forward filtering-backward smoothing algorithm can be implemented in a single forward pass in the case of \textit{additive functionals}, i.e. expectations of the form
\begin{equation}\label{add_func}
    \mathbb{E}_{p(x_{0:T}|y_{0:T})}\left[\sum_{t=0}^{T-1} g_t (x_t, x_{t+1})\right].
\end{equation}
Computing such expectations is useful for calibrating the state-space model parameters, see \cite{Poyiadjis2011, kantas2015, DelMoral2015}. The PaRIS algorithm \cite{Olsson2017} combines this technique with rejection sampling to implement an efficient and cheap version of forward filtering-backward simulation in a single forward pass (for additive functionals).
\par
Although it permits the implementation of these online algorithms, the requirement of additive functionals is very restrictive. Additionally, the forward only technique is \textit{function specific}, i.e. it directly updates an approximation to the expected value of the additive functional.
Our approach induces a controllable bias through the fixed-lag approximation but can approximate any expectation $\mathbb{E}_{p(x_{0:T}|y_{0:T})}[f(x_{0:T})]$ over the joint smoothing distribution and is therefore significantly more general than the marginal or additive functional approaches.
\par
Block sampling \cite{Doucet2006} is an online method that targets the joint smoothing distribution. Every time a new observation is received, the block sampling scheme discards the most recent coordinates (within lag $L$) and re-proposes from an enlarged proposal distribution based on many more recent observations. Although this scheme does make use of a fixed-lag parameter, the weights and resampling still act on the full joint smoothing distribution. By moving a larger proportion of the trajectories, block sampling can (when combined with adaptive resampling schemes) mitigate but not avoid path degeneracy as resampling still takes place over full trajectories.
Our proposed method builds on the block sampling approach by proposing blocks with a single coordinate overlap and then resampling in a way that targets the fixed-lag joint smoothing distribution \eqref{fixed_lag_j}.

\section{Fixed-lag Stitching}
\label{sec:stitch}

In this section, we first detail the technique underlying backward simulation and then describe how the same technique can be applied in a forward implementation under a fixed-lag approximation.
\par

\subsection{Backward Simulation}
\label{sec:back_sim}
In a single iteration of backward simulation, we have samples
\begin{align*}
    \left\{\tilde{x}_{t-1}^{(i)}, \tilde{w}_{t-1}^{(i)} \right\}_{i=1}^N &\text{ approximating } p(x_{t-1}|y_{0:t-1}), \\
    \left\{x_t^{(j)} \right\}_{j=1}^N &\text{ approximating } p(x_t|y_{0:T}),
\end{align*}
but desire samples from the joint
\begin{equation*}
    \left\{(x_{t-1}^{(j)}, x_t^{(j)} ) \right\}_{j=1}^N \text{ approximating } p(x_{t-1}, x_t|y_{0:T}).
\end{equation*}
Here, and in the next sections, we have used tildes to represent the particles from which we look to resample - particles without tildes remain fixed.
Seeing as $\{x_t^{(j)} \}_{j=1}^N$ already have the correct marginals this amounts to sampling
\begin{align*}
    x_{t-1}^{(j)} \sim p(x_{t-1}| x_t^{(j)}, y_{0:t-1}) && \text{for } j = 1,\dots, N.
\end{align*}
We cannot sample from $p(x_{t-1}| x_t^{(j)}, y_{0:t-1})$ directly, so we consider the decomposition
\begin{align*}
    p(x_{t-1}| x_t^{(j)}, y_{0:t-1}) &=
    \frac{
    p(x_t^{(j)}|x_{t-1}) p(x_{t-1}|y_{0:t-1})
    }{
    \int p(x_t^{(j)}|x_{t-1}) p(x_{t-1}|y_{0:t-1}) dx_{t-1}
    }.
\end{align*}
and then use $p(x_{t-1}|y_{0:t-1})$ to form an importance sampling approximation.
Specifically,
we use the particle filter's empirical approximation in the numerator and denominator to obtain the self-normalised weights \cite{Godsill2004, DelMoral2010}
\begin{equation*}
    w_{t-1}^{(i \leftarrow j)} = 
    \frac{\tilde{w}_{t-1}^{(i)}p(x_t^{(j)} | \tilde{x}_{t-1}^{(i)})}
    {\sum_{k=1}^N \tilde{w}_{t-1}^{(k)} p(x_t^{(j)} | \tilde{x}_{t-1}^{(k)})}.
\end{equation*}
where $\sum_{k=1}^N \tilde{w}_{t-1}^{(k)} p(x_t^{(j)} | \tilde{x}_{t-1}^{(k)})$ is an asymptotically unbiased approximation of $p(x_t^{(j)}|y_{0:t-1})$ and now $\sum_{i=1}^N w_{t-1}^{(i \leftarrow j)} = 1$.
\par
Sampling from the empirical distribution $\sum_{i=1}^N w_{t-1}^{(i \leftarrow j)} \delta_{x_{t-1}}(\tilde{x}_{t-1}^{(i)})$ directly for each $j$ is the $O(N^2)$ backward simulation technique \cite{Godsill2004}, which when iterated for $t=T,\dots, 1$ is asymptotically unbiased for the joint smoothing distribution $p(x_{0:T}|y_{0:T})$.

\subsection{Fixed-lag Forward Simulation - Intractable}\label{subsec:intrac_w}
\label{sec:fl_forward_sim_intract}
In this section we aim to approximate the fixed-lag joint smoothing distribution defined in \eqref{fixed_lag_j} during the forward pass of particle filtering thus obtaining an online algorithm.
In the setting of fixed-lag forward simulation we have
\begin{align*}
    \left\{x_{t-1}^{(i)} \right\}_{i=1}^N &\text{ approximating } p(x_{t-1}|y_{0:T-1}),\\
    \left\{\tilde{x}_t^{(j)}, \tilde{w}_{t}^{(j)} \right\}_{j=1}^N &\text{ approximating } p(x_t|y_{0:T}),
\end{align*}
where $t = T-L$. Recall the fixed-lag approximation uses all past, present and at most $L$ future observations to infer $x_{t-1}$, which implies  $p(x_{t-1}|y_{0:T-1}) \approx p(x_{t-1}|y_{0:T})$.
\par
We desire unweighted samples from the joint
\begin{align*}
    \left\{\left(x_{t-1}^{(i)}, x_t^{(i)} \right) \right\}_{i=1}^N \text{ approximating } p^L(x_{t-1}, x_t|y_{0:T}),
\end{align*}
where $p^L(x_{t-1}, x_t|y_{0:T}) = p(x_{t-1}|y_{0:T-1}) p(x_t|x_{t-1}, y_{t:T})$.
As $\left\{x_{t-1}^{(i)} \right\}_{i=1}^N$
have the correct marginal, obtaining the desired joint samples amounts to sampling
\begin{align*}
    x_{t}^{(i)} \sim p(x_{t}| x_{t-1}^{(i)}, y_{t:T}) && \text{for } i = 1,\dots, N.
\end{align*}
We cannot sample this conditional density directly and (similarly to backward simulation) employ importance sampling to approximate it.
\par
In order to use the empirical approximation of $p(x_t|y_{0:T})$ as the sampling density (for an online implementation) we need to write 
\[
 p(x_t|x_{t-1}^{(i)}, y_{t:T}) = \frac{h_t(x_{t-1}^{(i)},x_t)  p(x_t | y_{0:T})}
 {\int h_t(x_{t-1}^{(i)},x_t)  p(x_t | y_{0:T}) dx_t },
\]
for some suitably defined non-negative and integrable function $h_t(x_{t-1}^{(i)},x_t)$. We then use the set of samples $\left\{\tilde{x}_t^{(j)}, \tilde{w}_{t}^{(j)} \right\}_{j=1}^N$ in the numerator and also to approximate the denominator to 
obtain a set of normalised weights
\[
 w_{t}^{(i \rightarrow j)} = 
    \frac{ \tilde{w}_t^{(j)} h_t(x_{t-1}^{(i)}, \tilde{x}_t^{(j)} )}
{\sum_{k=1}^N  \tilde{w}_t^{(k)} h_t(x_{t-1}^{(i)},\tilde{x}_t^{(k)})}.
\]
To find $h_t$, 
\begin{align*}
    p(x_{t}| x_{t-1}^{(i)}, y_{t:T})
    &= p(x_{t}| x_{t-1}^{(i)}, y_{0:T}) \\
    &= \frac{
    p(y_{t:T}| x_t) p(x_t| x_{t-1}^{(i)})
    }{
    p(y_{t:T}|x_{t-1}^{(i)})
    }, \\
    &=
    \frac{
    p(y_{t:T}| x_t) p(x_t| x_{t-1}^{(i)})
    }{
    p(y_{t:T}|x_{t-1}^{(i)})
    p(x_t|y_{0:T})
    }
    p(x_t|y_{0:T}).
\end{align*}
where in the last line we have multiplied and divided by $p(x_t|y_{0:T})$. A further  application of Bayes' theorem to $p(x_t|y_{0:T})$ gives us
\begin{align*}
    p(x_{t}| x_{t-1}^{(i)}, y_{t:T}) =
    \frac{
    p(y_{t:T}| y_{0:t-1}) p(x_t| x_{t-1}^{(i)})
    }{
    p(y_{t:T}|x_{t-1}^{(i)})
    p(x_t|y_{0:t-1})
    }
    p(x_t|y_{0:T}).
\end{align*}
Since $ p(x_{t}| x_{t-1}^{(i)}, y_{t:T})$ integrates to 1, $h_t$ is clearly
\[
h_t(x_{t-1}^{(i)}, {x}_t ) = \frac{ p(x_t| x_{t-1}^{(i)})}{p(x_t|y_{0:t-1})}
\]
which is where we stop as the density $p(x_t|y_{0:t-1})$ is not tractable (note that we could replace $p(x_t|y_{0:t-1})$ with an empirical approximation using marginal filtering particles but the resulting algorithm would have a prohibitive $O(N^3)$ complexity).


\subsection{Fixed-lag Forward Simulation - Tractable}\label{subsec:trac_w}
\label{sec:fl_forward_sim_tract}
In the previous section, the $h_t(x_{t-1}^{(i)},x_t)$ was intractable. We address this problem by defining a new conditional density
\[
 \pi(x_{t}, x_{t-1}| x_{t-1}^{(i)}, y_{0:T}) =  p(x_t|x_{t-1}^{(i)}, y_{t:T}) \lambda({x}_{t-1} | x_{t-1}^{(i)}, x_t)
\]
which trivially admits $p(x_t|x_{t-1}^{(i)}, y_{t:T})$ as the marginal. The  potential dependency of $\lambda$ on $y_{0:T}$ is implicit. The conditional density $\lambda$ is to be chosen and we show how to make this choice so that when using $p({x}_{t-1}, x_t|y_{0:T})$ within an importance sampling approximation of $\pi$, the weight is now tractable. In keeping with the notation of the previous section,
\begin{align*}
    \left\{ (\tilde{x}_{t-1}^{(j)}, \tilde{x}_t^{(j)}) , \tilde{w}_{t}^{(j)} \right\}_{j=1}^N &\text{ approximates } p({x}_{t-1}, x_t|y_{0:T}),
\end{align*}
where we have used tildes to represent the particles from which we look to resample - non-tilde particles are fixed. We remark that this idea of resolving the intractability of weights via sampling a higher dimensional density is inspired by the work of \cite{Doucet2006} for blocked resampling of the path-space particle approximations.

\begin{proposition}
Let $ \lambda({x}_{t-1} | x_{t-1}^{(i)}, x_t) = p(x_{t-1} | y_{0:t-1})$ and $H(x_{t-1}^{(i)},x_{t-1}, x_t)=p(x_t | x_{t-1}^{(i)}) / p(x_t | x_{t-1})$ then 
\begin{equation*}
 \pi(x_{t}, x_{t-1}| x_{t-1}^{(i)}, y_{0:T})
  =\frac{H(x_{t-1}^{(i)},x_{t-1}, x_t) p(x_{t}, x_{t-1} | y_{0:T}) }{\int H(x_{t-1}^{(i)},x_{t-1}, x_t) p(x_{t}, x_{t-1}|  y_{0:T}) dx_{t-1:t}}.
\end{equation*}
The self-normalised approximation of $p(x_{t} | x_{t-1}^{(i)}, y_{0:T})$ is 
$ \left\{ \tilde{x}_t^{(j)} , \tilde{w}_{t}^{(i \rightarrow j)} \right\}$
where
\begin{align}\label{stitch_weights}
    w_t^{(i \rightarrow j)} = \tilde{w}_{t}^{(j)} \frac{
    p(\tilde{x}_t^{(j)}|{x}_{t-1}^{(i)})
    }{
    p(\tilde{x}_t^{(j)}|\tilde{x}_{t-1}^{(j)})
    }
    \left(
    \sum_{k=1}^N
    \tilde{w}_{t}^{(k)}
    \frac{
    p(\tilde{x}_t^{(k)}|{x}_{t-1}^{(i)})
    }{
    p(\tilde{x}_t^{(k)}|\tilde{x}_{t-1}^{(k)})
    }
    \right)^{-1}.
\end{align}
\end{proposition}
The interpretation of this proposition for forward smoothing is as follows: we can now sample from this approximation of $\pi$ directly. Discarding the sampled ${x}_{t-1}$ leaves samples $(x_{t-1}^{(i)}, x_{t}^{(i)})$ from the desired joint $p^L(x_{t-1}, x_{t}|y_{0:T})$. Repeating this for each $i$ results in an $O(N^2)$ algorithm that is asymptotically unbiased for the fixed-lag joint distribution.
\par

\begin{proof}
Expanding the density $\pi$ yields
\begin{align*}
    \pi(x_{t}, {x}_{t-1}| x_{t-1}^{(i)}, y_{0:T})
    &=
    p(x_t|x_{t-1}^{(i)}, y_{t:T})\lambda({x}_{t-1} | x_{t-1}^{(i)}, x_t), \\
    &=
    \frac{
    p(y_{t:T}| x_t) p(x_t|x_{t-1}^{(i)})
    }{
    p(y_{t:T}|x_{t-1}^{(i)})
    }
    \lambda({x}_{t-1} | x_{t-1}^{(i)}, x_t).
\end{align*}
Dividing by the sampling distribution $p({x}_{t-1}, x_t|y_{0:T}) = p({x}_{t-1}|y_{0:T})p(x_t |{x}_{t-1}, y_{t:T})$  gives
\begin{equation*}
    \pi(x_{t}, {x}_{t-1}| x_{t-1}^{(i)}, y_{0:T}) / p({x}_{t-1}, x_t|y_{0:T})
    =
    \frac{
    p(y_{t:T}| x_t) p(x_t|x_{t-1}^{(i)})
    }{
    p(y_{t:T}|x_{t-1}^{(i)}) p({x}_{t-1}, x_t|y_{0:T})
    }
    \lambda({x}_{t-1} | x_{t-1}^{(i)}, x_t).
    \end{equation*}
Bayes' theorem on $p({x}_{t-1}, x_t|y_{0:T})$ gives
\begin{equation*}
    \pi(x_{t}, {x}_{t-1}| x_{t-1}^{(i)}, y_{0:T}) / p({x}_{t-1}, x_t|y_{0:T})=
    \frac{
    p(x_t|x_{t-1}^{(i)}) p(y_{t:T}|y_{0:t-1})
    }{
    p(x_t|{x}_{t-1}) p(y_{t:T}|x_{t-1}^{(i)}) p({x}_{t-1}|y_{0:t-1})
    }
    \lambda({x}_{t-1} | x_{t-1}^{(i)}, x_t).
\end{equation*}
We now observe that the choice of $\lambda({x}_{t-1} | x_{t-1}^{(i)}, x_t,) = p({x}_{t-1}|y_{0:t-1})$ will make all terms involving $({x}_{t-1}, x_t)$ tractable.
\begin{equation*}
    \pi(x_{t}, {x}_{t-1}| x_{t-1}^{(i)}, y_{0:T})
    =
    \frac{
    p(x_t|x_{t-1}^{(i)}) p(y_{t:T}|y_{0:t-1})
    }{
    p(x_t| {x}_{t-1}) p(y_{t:T}|x_{t-1}^{(i)})
    } p({x}_{t-1}, x_t|y_{0:T}).
\end{equation*}
\end{proof}

\begin{algorithm}
	\caption{Fixed-lag Particle Stitching}
	\begin{algorithmic}[1]\label{alg:hybrid_stitch}
		\item Input unweighted sample $\{ x_{0:T-L-1}^{(i)} \}_{i=1}^N$, weighted sample $\{ \tilde{x}_{T-L-1:T}^{(j)}, \tilde{w}_T^{(j)} \}_{j=1}^N$, bound $\rho \geq p(x_{T-L}| x_{T-L-1})$ and the maximum number of rejections to attempt $R$.
		\STATE Calculate the non-interacting stitching weights and normalise in $j$
		\begin{align*}
			\hat{w}_T^{(j)} &\propto
			\frac{1}
			{p(\tilde{x}_{T-L}^{(j)}|\tilde{x}_{T-L-1}^{(j)})}
			\tilde{w}_T^{(j)}
			\tag*{$j=1,\dots,N$.}
		\end{align*}
		\FOR{$i=1,\dots,N$}
		\FOR{$r=1,\dots,R$}
		\STATE Sample $c^* \sim \text{Categorical} \left( \left\{ \hat{w}_T^{(j)} \right\}_{j=1}^N \right)$
		\STATE Sample $u \sim U(0,1)$
		\IF{$u < p(\tilde{x}_{T-L}^{(c^*)}|x_{T-L-1}^{(i)})/\rho$}
		\STATE Accept $c^*$
		\STATE \textbf{break}
		\ENDIF
		\ENDFOR
		\IF{a sample $c^*$ was accepted}
		\STATE Set $x_{T-L:T}^{(i)} = \tilde{x}_{T-L:T}^{(c^*)}$
		\ELSE
		\STATE Calculate the stitching weights \label{alg:step:calc_norm_sw} and normalise in $j$
		\begin{align*}
			w_T^{(i\rightarrow j)} &\propto
			\frac{p(\tilde{x}_{T-L}^{(j)}|x_{T-L-1}^{(i)})}
			{p(\tilde{x}_{T-L}^{(j)}|\tilde{x}_{T-L-1}^{(j)})}
			\tilde{w}_T^{(j)}
			\tag*{$j=1,\dots,N$.}
		\end{align*}
		\STATE Sample $c_i \sim \text{Categorical} \left( \left\{ w_T^{(i\rightarrow j)} \right\}_{j=1}^N \right)$
		\STATE Set $x_{T-L:T}^{(i)} = \tilde{x}_{T-L:T}^{(c_i)}$ \label{alg:step:samp_sw}
		\ENDIF
		\ENDFOR
		\STATE Output unweighted sample $\left\{x_{0:T}^{(i)}\right\}_{i=1}^N$ approximating $p^L(x_{0:T}|y_{0:T})$.
	\end{algorithmic}
\end{algorithm}

In \eqref{stitch_weights}, we have described above an empirical approximation that stitches together particles from $p(x_{T-L-1} |y_{0:T-1})$ with those from $p(x_{T-L-1:T-L} | y_{0:T})$.
By the conditional independence structure of state-space models this is equivalent to stitching together blocks from $p^L(x_{0:T-L-1} |y_{0:T-1})$ with those from $p(x_{T-L-1:T} | y_{0:T})$ - assuming we can sample from $p(x_{T-L-1:T} | y_{0:T})$.
\par
Sampling from $p(x_{T-L-1:T}|y_{0:T})$ is not directly possible for non-trivial state-space models. In Section~\ref{sec:block_prop} we describe two efficient methods for block sampling by recycling previously generated particles (Alg.~\ref{alg:joint_fl_pf} and Alg.~\ref{alg:joint_fl_ffbsi}).
\par
The fixed-lag stitching described provides a method to, in principal, approximate the full smoothing distribution $p(x_{0:T} | y_{0:T})$. It does so by using very separate tools to the forward-only techniques in \cite{DelMoral2010, Olsson2017}, these techniques do not require any stitching and directly update an expectation over additive functionals. Our method generates a collection of particles approximating the joint smoothing distribution and is therefore far more general. We have utilised the block sampling framework from \cite{Doucet2006}, in particular through the balancing distribution $\lambda$. However, the fixed-lag stitching is novel and effectively uses a fixed-lag approximation to implement a forward implementation of the technique underlying backward simulation \cite{Godsill2004}. This is in contrast to \cite{Doucet2006}, where particles are reweighted and resampled over full trajectories and therefore still suffer from path degeneracy.

\subsection{Rejection Sampling}

Sampling from \eqref{stitch_weights} can be done directly at a computational complexity of $O(N^2)$. However, when a bound for the transition density is available
\begin{equation}\label{transition_bound}
    \rho \geq p(x_{T-L}| x_{T-L-1}) \quad \forall x_{T-L-1}, x_{T-L},
\end{equation}
we can utilise the rejection sampling approach of \cite{Douc2011} to avoid calculating the normalisation constants and bring the computational complexity down to $O(N)$.
In practice, the rejection sampling is not always faster than the direct one. A pragmatic approach is the hybrid described in \cite{Taghavi2013} where for each particle, up to $R<N$ samples are proposed to the rejection sampler, if all are subsequently rejected the direct scheme is applied, thus setting $R=0$ recovers the direct scheme. This hybrid algorithm is described in Alg.~\ref{alg:hybrid_stitch}.
\par

\section{Sampling from $p(x_{T-L-1:T}|y_{0:T})$}
\label{sec:block_prop}
We now describe two methods for sampling coordinates $\tilde{x}_{T-L-1:T}$ in a way that is asymptotically unbiased for $p(x_{T-L-1:T}|y_{0:T})$, and can therefore be plugged into the fixed-lag stitching procedure. \par

\subsection{Particle Filter}
Recall the online setting where we have unweighted particles $\{ x_{0:T-1}^{(i)} \}_{i=1}^N$ approximating $p^L(x_{0:T-1}|y_{0:T-1})$ and receive a new observation $y_T$.
\par
Our first method is based on the fact that the particle approximation provided by a classical particle filter is asymptotically unbiased for the full joint smoothing distribution $p(x_{0:T}|y_{0:T})$. Although this approximation deteriorates due to path degeneracy it may still be sufficient for sampling the recent coordinates $x_{T-L-1:T}$.
\par
Thus, we propose applying the classical particle filter proposal and weighting steps to $\{ x_{0:T-1}^{(i)} \}_{i=1}^N$, generating weighted particles $\{  x_{0:T}^{(i)}, \tilde{w}_T^{(i)}  \}_{i=1}^N$, before splitting the trajectories
\begin{equation*}
    \left\{ x_{0:T}^{(i)}, \tilde{w}_T^{(i)} \right\}_{i=1}^N
    \rightarrow
    \left\{ x_{0:T-L-1}^{(i)} \right\}_{i=1}^N
    \text{and}
    \left\{ \tilde{x}_{T-L-1:T}^{(i)}, \tilde{w}_T^{(i)} \right\}_{i=1}^N,
\end{equation*}
where $\tilde{x}_{T-L-1:T}^{(i)} = x_{T-L-1:T}^{(i)}$.
Under the fixed-lag approximation, $x_{0:T-L-1}$ is conditionally independent of $y_T$ and therefore the new weights need not apply to these earlier coordinates. Whereas the $\{ \tilde{x}_{T-L-1:T}^{(i)}, \tilde{w}_T^{(i)} \}_{i=1}^N$ are asymptotically unbiased for the desired sampling distribution $p(x_{T-L-1:T}|y_{0:T})$ and can therefore be plugged into the stitching procedure, Alg.~\ref{alg:hybrid_stitch}. The coordinates $x_{T-L-1}$ are duplicated to provide the overlap required for stitching.
\par

\begin{algorithm}
	\caption{Online Particle Smoother (for $T>L$)}
	\begin{algorithmic}[1]  \label{alg:joint_fl_pf}
	    \item Input unweighted smoothing sample $\{ x_{0:T-1}^{(i)} \}_{i=1}^N$, new observation $y_T$ and fixed-lag $L$.
	    \STATE Fix $\{ x_{0:T-L-1}^{(i)} \}_{i=1}^N$
	    \STATE Execute particle filter propagate and weight steps, Alg. \ref{alg:pf}: steps \ref{alg:step:pf:prop}:\ref{alg:step:pf:weight}
	    \begin{multline*}
	        \text{Generate } \{ \tilde{x}_{T-L-1:T}^{(j)}, \tilde{w}_T^{(j)} \}_{j=1}^N \\
	        \text{ from } \{ x_{T-L-1:T-1}^{(j)} \}_{j=1}^N
	        \text{ and } y_T,
	    \end{multline*}
	    forming a weighted sample approximating $p(x_{T-L-1:T}|y_{0:T})$.
	    \STATE Stitch together
	    \begin{align*}
	        \{ x_{0:T-L-1}^{(i)} \}_{i=1}^N \text{ and } \{ \tilde{x}_{T-L-1:T}^{(j)}, \tilde{w}_T^{(j)}\}_{j=1}^N
	        \rightarrow 
	        \{ x_{0:T}^{(i)} \}_{i=1}^N.
	    \end{align*}
	    using Alg. \ref{alg:hybrid_stitch}.
	    \STATE Output unweighted sample $\{ x_{0:T}^{(i)} \}_{i=1}^N$ approximating $p^L(x_{0:T}|y_{0:T})$
	\end{algorithmic}
\end{algorithm}

The algorithm is described in Alg.~\ref{alg:joint_fl_pf}, and ends up being a relatively simple modification to a classical particle filter where the resampling step is compulsory and altered to include the stitching probabilities in the weights \eqref{stitch_weights}.
\par
When the transition bound \eqref{transition_bound} is available the complexity of the update remains $O(N)$ or $O(N^2)$ when the bound is unavailable.

\subsection{Partial Backward Simulation}
If the lag parameter $L$ is chosen to be too large, the above mechanism will still suffer from path degeneracy in the same way a particle filter does. To remedy this we propose a partial run of backward simulation (Alg.~\ref{alg:ffbsi}) at each time step to rejuvenate the trajectories $x_{T-L-1:T}$. This technique is considered in \cite{Clapp1999} for generating samples from $p(x_{T-L:T}|y_{0:T})$ without the subsequent stitching.
\par
This has the additional requirement of storing the marginal approximations $\{\tilde{x}_t^{(k)}, \tilde{w}_t^{(k)} \}_{k=1}^N$ for $t = T{-}L{-}1,\dots,T$ from the particle filter, but permits the use of adaptive resampling and completely avoids path degeneracy.
\par
The resulting algorithm, Alg.~\ref{alg:joint_fl_ffbsi}, has a complexity of $O(LN)$ per update if the transition bound \eqref{transition_bound} is available otherwise $O(LN^2)$.
\par

\begin{algorithm}
	\caption{Online Particle Smoother with Backward Simulation (for $T>L$)}
	\begin{algorithmic}[1]  \label{alg:joint_fl_ffbsi}
	    \item Input unweighted smoothing sample $\{ x_{0:T-1}^{(i)} \}_{i=1}^N$, weighted marginal filtering samples $\{ \tilde{x}_{t}^{(k)}, \tilde{w}_t^{(k)}\}_{k=1}^N$ for $t=T{-}L{-}1,\dots,T-1$, new observation $y_T$, and fixed-lag $L$.
	    \STATE Fix $\{ x_{0:T-L-1}^{(i)} \}_{i=1}^N$
	    \STATE Execute particle filter update, Alg. \ref{alg:pf}, to generate the new marginal filtering sample
	    \begin{multline*}
	        \text{Generate } \{ \tilde{x}_{T}^{(k)}, \tilde{w}_T^{(k)} \}_{k=1}^N \\
	        \text{ from } \{ \tilde{x}_{T-1}^{(k)}, \tilde{w}_{T-1}^{(k)} \}_{k=1}^N
	        \text{ and } y_T.
	    \end{multline*}
	    \STATE Run partial backward simulation, Alg.~\ref{alg:ffbsi}, on the weighted filtering samples
	    \begin{multline*}
	        \{ \tilde{x}_{t}^{(k)}, \tilde{w}_t^{(k)}\}_{k=1}^N, \text{ for } t=T,\dots,T{-}L{-}1
	        \\
	        \rightarrow
	        \{ \tilde{x}_{T-L-1:T}^{(j)} \}_{j=1}^N.
	    \end{multline*}
	    forming an unweighted sample approximating $p(x_{T-L-1:T}|y_{0:T})$.
	    \STATE Stitch together
	    \begin{align*}
	        \{ x_{0:T-L-1}^{(i)} \}_{i=1}^N \text{ and } \{ \tilde{x}_{T-L-1:T}^{(j)}, \tfrac1N\}_{j=1}^N
	        \rightarrow 
	        \{ x_{0:T}^{(i)} \}_{i=1}^N.
	    \end{align*}
	    using Alg. \ref{alg:hybrid_stitch}.
	    \STATE Output unweighted sample $\{ x_{0:T}^{(i)} \}_{i=1}^N$ approximating $p^L(x_{0:T}|y_{0:T})$ and weighted filtering samples $\{ \tilde{x}_{T}^{(k)}, \tilde{w}_T^{(k)}\}_{k=1}^N$, $t=T-L,\dots,T$.
	\end{algorithmic}
\end{algorithm}

Both techniques to sample from $p(x_{T-L-1:T}|y_{0:T})$ utilise the output of a particle filter. Indeed, they are also applicable to alternative filtering techniques such as auxiliary particle filters \cite{Pitt1999} and the backward simulation approach is also applicable to filters that discard historic trajectories such as the marginal particle filter \cite{Klaas2005}.

\section{Map-Matching}
\label{sec:mm}

\begin{figure}
    \centering
    \begin{subfigure}{0.48\columnwidth}
    \centering
        \includegraphics[width=\textwidth]{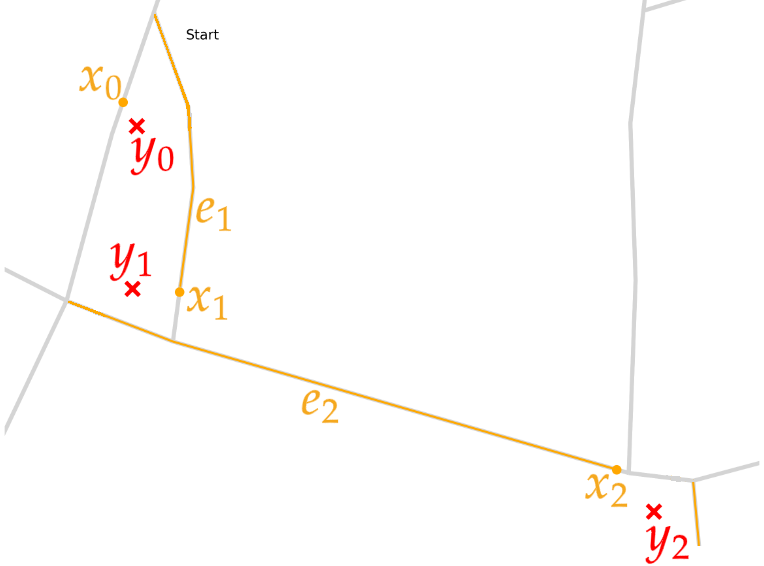}
    \caption{Arbitrary stitching, \cite{Kitagawa2001}, Alg.~\ref{alg:partial_res}. Discontinuities where the vehicle jumps, e.g. $e(x_0) \neq e^o_1$.}
    \label{fig:arb_stitch}
    \end{subfigure}
    \begin{subfigure}{0.48\columnwidth}
    \centering
        \includegraphics[width=\textwidth]{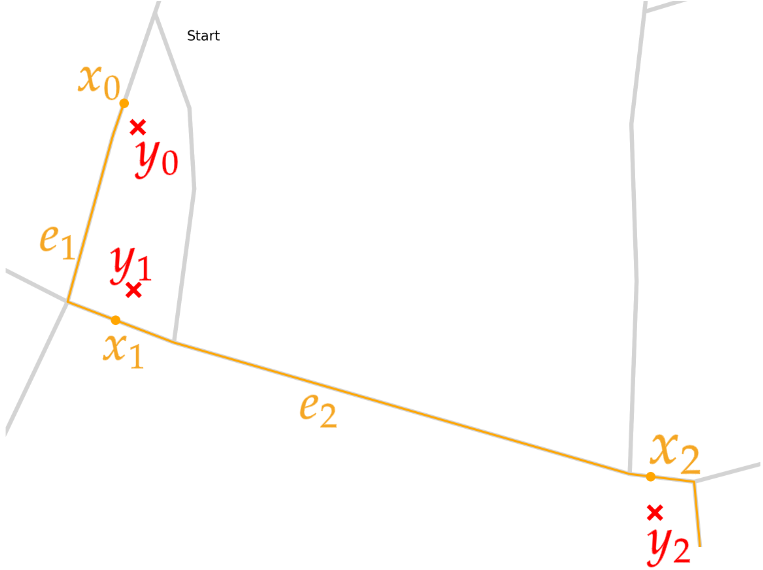}
        \caption{Fixed-lag stitching, Alg.~\ref{alg:hybrid_stitch} \\}
        \label{fig:stitch}
    \end{subfigure}
    \caption{Comparison of fixed-lag techniques for map-matching. Each image displays a single trajectory from the resulting particle approximation. Both techniques produce plausible marginals but arbitrary stitching fails to produce a continuous trajectory.}
    \label{fig:arb_vs_stitch}
\end{figure}

Map-matching is the task of inferring the true trajectory of a vehicle given noisy GPS observations and a map of the road network. A road network is defined as a graph within $\mathbb{R}^2$, where intersections are represented by nodes (vertices) and roads (assumed to be single lanes and one-way) are represented by edges with a two-way road being represented by two edges. Some collections of nodes and edges are depicted in \figurename~\ref{fig:arb_vs_stitch}, \ref{fig:synth_route} and \ref{fig:single_start_route}, with map-matched vehicle routes overlaid. The recorded GPS observations (see  \figurename~\ref{fig:arb_vs_stitch}) may lie outside the road whereas valid vehicle positions/trajectories to be inferred must strictly lie on the road.
Following \cite{Newson2009, Roth2012} for urban road networks we infer the vehicle's position along an edge but not it's width within the road.
\par


Applications of map-matching are wide-ranging and thus a general purpose algorithm is highly desirable. Existing probabilistic approaches to map-matching have mostly adopted the approach of \cite{Newson2009}, where each observation is snapped to the nearest point on any and all edges that fall within a truncation distance. These points then form a discrete hidden Markov model, on which the Viterbi algorithm can produce a single route of high probability. In contrast, our approach provides uncertainty quantification through a collection of particles, with each particle representing a possible route. Previous applications of particle filtering to map-matching \cite{Davidson2011, Kempinkska2016} fail to tackle the problem of path degeneracy, with the exception of \cite{Roth2012} who introduce the use of FFBSi for offline map-matching. Our formulation is similar to that of \cite{Roth2012} but differs through the inclusion of a term in the transition density adapted from \cite{Newson2009} that penalises non-direct routes (which is vital in dense urban road networks), as well as the use of the optimal proposal density (that takes into account the new observation) rather than simply the bootstrap proposal which will perform poorly for small GPS noise or dense road networks.
\par

Traditional fixed-lag smoothing \cite{Kitagawa2001}, Alg.~\ref{alg:partial_res} where particles are arbitrarily stitched together is not appropriate for map-matching, as seen in \figureautorefname~\ref{fig:arb_vs_stitch}. As these methods target the product of smoothing marginals in \eqref{fixed_lag_m}, arbitrary stitching cannot produce a continuous trajectory and therefore does not permit expectations over multiple observation times.
\par

In the rest of this section, we describe our state-space model for map-matching, the optimal proposal distribution and the induced weights in the context of fixed-lag stitching. In the next section we present some results on synthetic and real data.
An open source python package providing easy offline and online map-matching is provided \footnote{\url{https://github.com/SamDuffield/bmm}} alongside code for all the simulations to follow.

\subsection{Model Variables} \label{model-var}
\begin{figure}
	\centering
	\includegraphics[width=0.6\columnwidth]{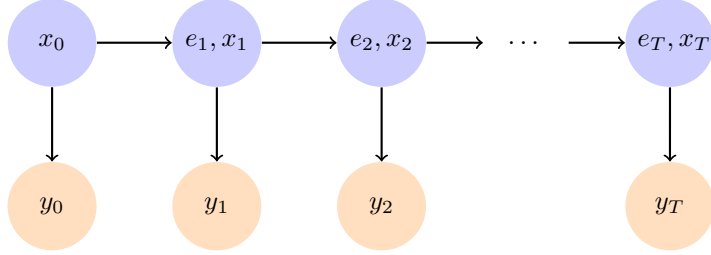}
	\caption{Conditional independence structure of the state-space model for map-matching.}
	\label{fig:mm_ssm}
\end{figure}

We define a state-space model for a single vehicle's trajectory with the variables
\begin{itemize}
    \item $e_t \subset \mathbb{N}$, a finite ordered set of edge labels that define a connected path. Each edge label corresponds to a unique one-way section of road. The variable $e_t$ details the edges traversed (and in which order) between observation times $t{-}1$ and $t$, including the choices made at encountered intersections (nodes).
    \item $x_t \in \mathbb{R}^2$ the position of the vehicle at observation time. The variable $x_t$ defines a Cartesian coordinate restricted to lie on the road network, specifically $x_t$ lies on the final edge of the finite ordered set of edge labels  $e_t$, for $t>0$. 
    \item $y_t \in \mathbb{R}^2$ noisy observation of the vehicle's position $x_t$, not restricted to the road network.
\end{itemize}
Note here the change of notation from the previous section, now $x_t$ refers only to vehicle position and the full latent states are $x_0, (e_1, x_1), (e_2, x_2), \dots, (e_T, x_T)$ as depicted in \figurename~\ref{fig:mm_ssm}.
\par
We denote $e_t^o$ for the first edge in the ordered set $e_t$ and $e_t^*$ for the final edge, thus $e_t=(e_t^o,\ldots,e_t^*)$. We use the notation $e(x_t)$ to denote the edge label on which $x_t$ lies, so  $e(x_{t-1}) = e_t^o$ and $e(x_t) = e_t^*$.

\subsection{Model Distributions} \label{model-dens}
Our transition density can be written as
\begin{align}
    p(e_t, x_t | x_{t-1}) =
    \frac{
    \gamma(\|x_t - x_{t-1} \|_{e_t}) 
    \exp(-\beta | \|x_t - x_{t-1} \|_{e_t} - \|x_t - x_{t-1} \| |)}{
    Z(x_{t-1})},
     \label{eq:trans_dens}
\end{align}
with normalising constant
\begin{align}  \label{eq:norm_trans_dens} 
    Z(x_{t-1}) = \sum_{e_t} \int_{x_t} \gamma(\|x_t - x_{t-1} \|_{e_t}) 
    \exp(-\beta | \|x_t - x_{t-1} \|_{e_t} - \|x_t - x_{t-1} \| |) \textrm{d} x_t.
\end{align}
The summation in \eqref{eq:norm_trans_dens} is taken over all possible series of edges starting at $e(x_{t-1})$.
$\|x_t - x_{t-1} \|_{e_t}$ is the distance travelled between $x_{t-1}$ and $x_t$ along the series $e_t$ (restricted to the road network) whereas $\|x_t - x_{t-1} \|$ is the \textit{great circle} distance (not restricted to the road network).
\par
Thus the following distributions fully define our state-space model:
\begin{itemize}
    \item $\gamma(\|x_t - x_{t-1} \|_{e_t})$. Prior on distance travelled between observations - some simple analytical distribution on $\mathbb{R}^+$, penalising lengthy routes. We assume an exponential distribution with probability mass at 0 to represent the possibility of the vehicle remaining stationary (due to traffic lights, heavy traffic etc)
    \begin{align}
    \gamma(d) = p^0 \mathbb{I}(d = 0)  \label{mm_distance_prior}
    + (1{-}p^0) \mathbb{I}(d>0) \lambda e^{-\lambda d}.
	\end{align}
    \item $\exp(-\beta | \|x_t - x_{t-1} \|_{e_t} - \|x_t - x_{t-1} \| |)$ adapted from \cite{Newson2009}, penalising non-direct (or winding) routes. Non-direct routes with lots of curvature will have a high discrepancy between the road distance travelled and great circle distance and thus will have a low probability under this term, reflecting a driver's preference to take short, direct routes where possible.
    \item $p(y_t | x_t) = \mathcal{N}(y_t | x_t, \sigma^2_\text{GPS} I_2)$. Isotropic Gaussian observation noise.
    \item We set $p(x_0)$ to be uniform on the road network. I.e. no prior information on the start of the trajectory other than constricting it to the road network (as with all inferred positions).
\end{itemize}
To make $p(x_0 | y_0)$ tractable we define the initial observation density to be a truncated Gaussian:
\begin{equation*}
    p(y_0 | x_0) \propto \mathcal{N}(y_0 | x_0, \sigma^2_\text{GPS} I_2) \: \mathbb{I} (|| y_0 - x_0|| < r_\text{GPS}),
\end{equation*}
giving
\begin{equation*}
   p(x_0 | y_0) \propto \mathcal{N}(x_0 | y_0, \sigma^2_\text{GPS} I_2) \:
    \mathbb{I} (||y_0 - x_0|| < r_\text{GPS}),
\end{equation*}
where $x_0$ is restricted to the road network but $y_0$ is not. In our simulations we set $r_\text{GPS} = 5\sigma_\text{GPS}$.

\subsection{Optimal Proposal} \label{subsec:opt_prop}
The (locally) optimal proposal \cite{Doucet2000} for particle filtering combines the transition density \eqref{eq:trans_dens} and the newly received observation $y_T$:
\begin{align}
    q^\text{opt}\left(x_T, e_T| x_{T-1}, y_T \right)
    & \propto
    p(e_T, x_T| x_{T-1})
    p(y_T | x_T).
    \label{eq:opt_prop}
\end{align}
The standard reweighting step of the particle filter update (Alg. \ref{alg:pf}) then becomes
\begin{equation*}
    w^{\text{opt} \; (i)}_T \propto p(y_T | x^{(i)}_{T-1}),
\end{equation*}
where
\begin{equation}
    p(y_T | x_{T-1}) =  \sum_{e_T} \int_{x_T}
    p(e_T, x_T| x_{T-1})
    p(y_T | x_T)\
    \textrm{d} x_T,
    \label{eq:opt_int}
\end{equation}
is the normalisation constant of \eqref{eq:opt_prop}. \par

Neither sampling from the optimal proposal \eqref{eq:opt_prop}, nor evaluating the subsequent weights \eqref{eq:opt_int}, nor evaluating the normalising constant of the transition density \eqref{eq:norm_trans_dens} are immediately tractable as we do not have closed form expressions for the edge geometries. \par
Instead, we opt to approximate the required integrals numerically by discretising the edges up to some maximal possible distance travelled $d_{\text{max}}$. This numerical integration can be implemented efficiently across particles by caching route searches and likelihood evaluations, as many particles will typically lie on the same or adjacent edges.

\subsection{Fixed-lag Stitching}

In the context of the fixed-lag stitching described in Section~\ref{sec:stitch}, we propose stitching together each
\begin{equation*}
    (x_{0:T-L-1}^{(i)}, e_{1:T-L-1}^{(i)})
    \text{ with a sample from }  \left\{ (\tilde{x}_{T-L-1:T}^{(j)}, \tilde{e}_{T-L:T}^{(j)}) \right\}_{j=1}^N.
\end{equation*}
Thus the adjusted weights become 
\begin{align*}
    w^{(i\rightarrow j)}_T &\propto
    \frac{p(\tilde{e}_{T-L}^{(j)}, \tilde{x}_{T-L}^{(j)} | x_{T-L-1}^{(i)})}
    {p(\tilde{e}_{T-L}^{(j)}, \tilde{x}_{T-L}^{(j)} | \tilde{x}_{T-L-1}^{(j)})} w_T^{(j)} \mathbb{I}\left(e(x_{T-L-1}^{(i)}) =  \tilde{e}_{T-L}^{o \;(j)}\right).
\end{align*}

Similarly, for the rejection sampling we get non-interacting weights
\begin{equation*}
    \hat{w}_T^{(j)} \propto \frac{w^{(j)}_T}
	                    {p(\tilde{e}_{T-L}^{(j)}, \tilde{x}_{T-L}^{(j)} | \tilde{x}_{T-L-1}^{(j)})},
\end{equation*}
and we accept a sample if $e(x_{T-L-1}^{(i)}) =  \tilde{e}_{T-L}^{o \;(j)}$ and
\begin{align*}
    u < &\gamma \left(\| \tilde{x}_{T-L}^{(j)} - x_{T-L-1}^{(i)} \|_{\tilde{e}_{T-L}^{(j)}} \right)
    \\
    &\exp \left( -\beta \left| \| \tilde{x}_{T-L}^{(j)} - x_{T-L-1}^{(i)} \|_{\tilde{e}_{T-L}^{(j)}} -\| \tilde{x}_{T-L}^{(j)} - x_{T-L-1}^{(i)} \|\right| \right) 
    \\
     &\rho^{-1},
\end{align*}
where $u \sim U(0,1)$ and we have a bound $\rho > \gamma \left(d \right)$ for any $d$. The availability of this bound depends on the choice of distribution for $\gamma(\cdot)$, in the case of \eqref{mm_distance_prior} a bound is available: $\rho = \max((1-p^0)\lambda, p^0)$.


\section{Simulations}
\label{sec:results}

\begin{figure}
	\centering
	\begin{subfigure}{0.48\columnwidth}
		\centering
		\includegraphics[width=0.5\textwidth]{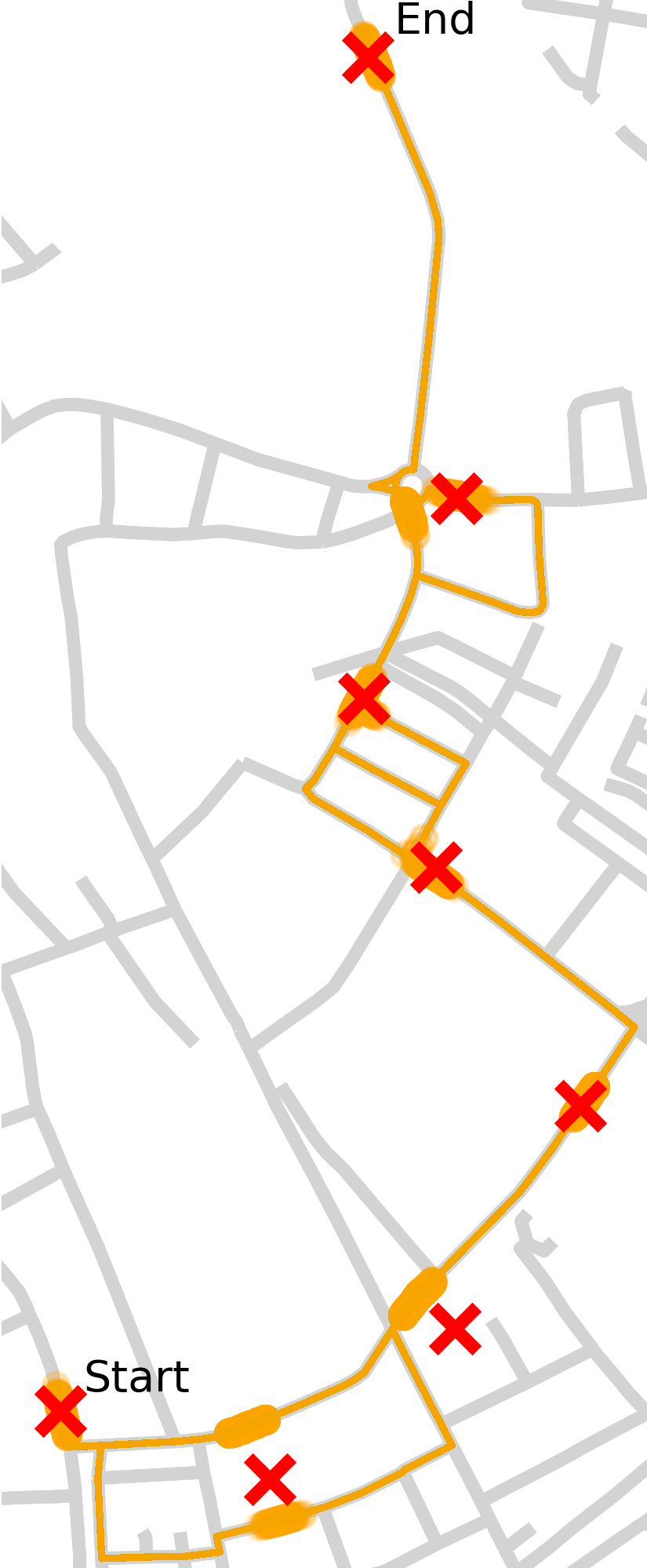}
		\caption{Gold standard, offline FFBSi with $N=1000$. All trajectories overlaid.}
		\label{fig:synth_ffbsi}
	\end{subfigure}
	\begin{subfigure}{0.48\columnwidth}
		\centering
		\includegraphics[width=0.5\textwidth]{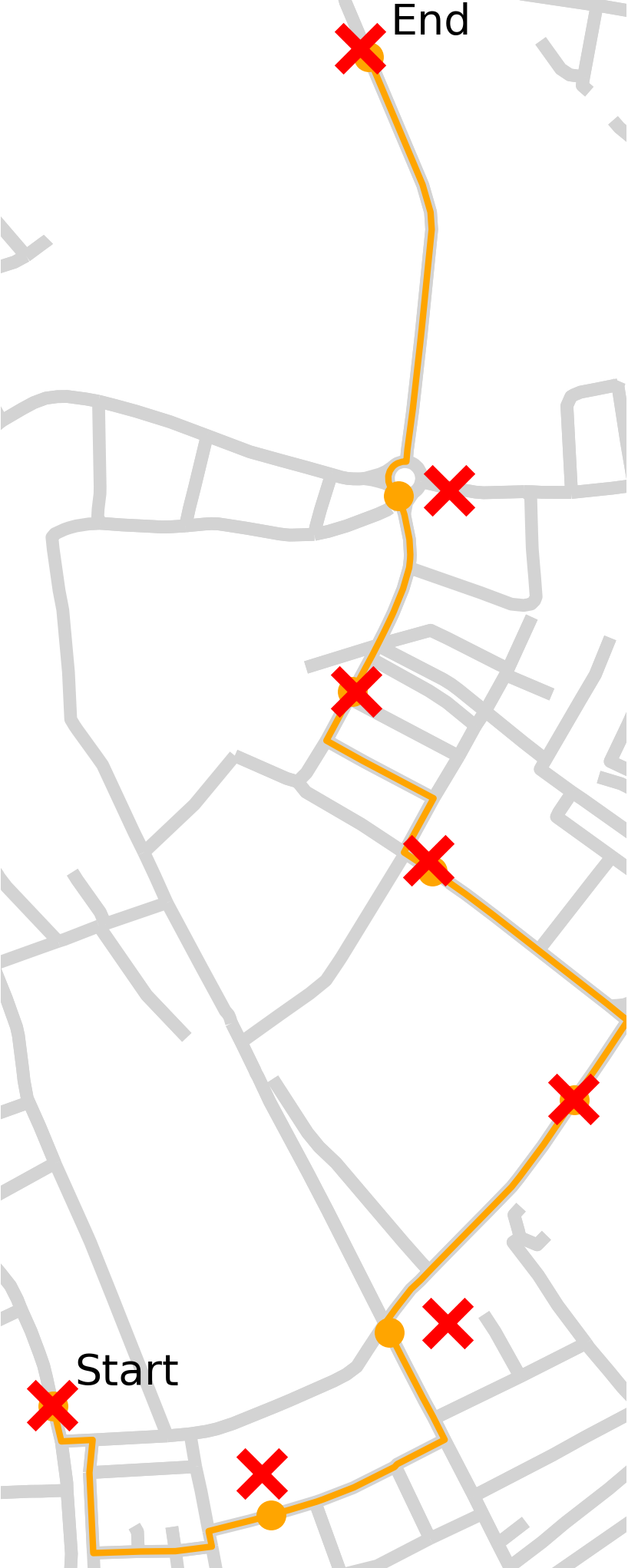}
		\caption{Viterbi map-matching \cite{Newson2009}. \newline \newline}
		\label{fig:synth_optim}
	\end{subfigure}
	\begin{subfigure}{0.7\columnwidth}
		\centering
		\includegraphics[width=0.7\textwidth]{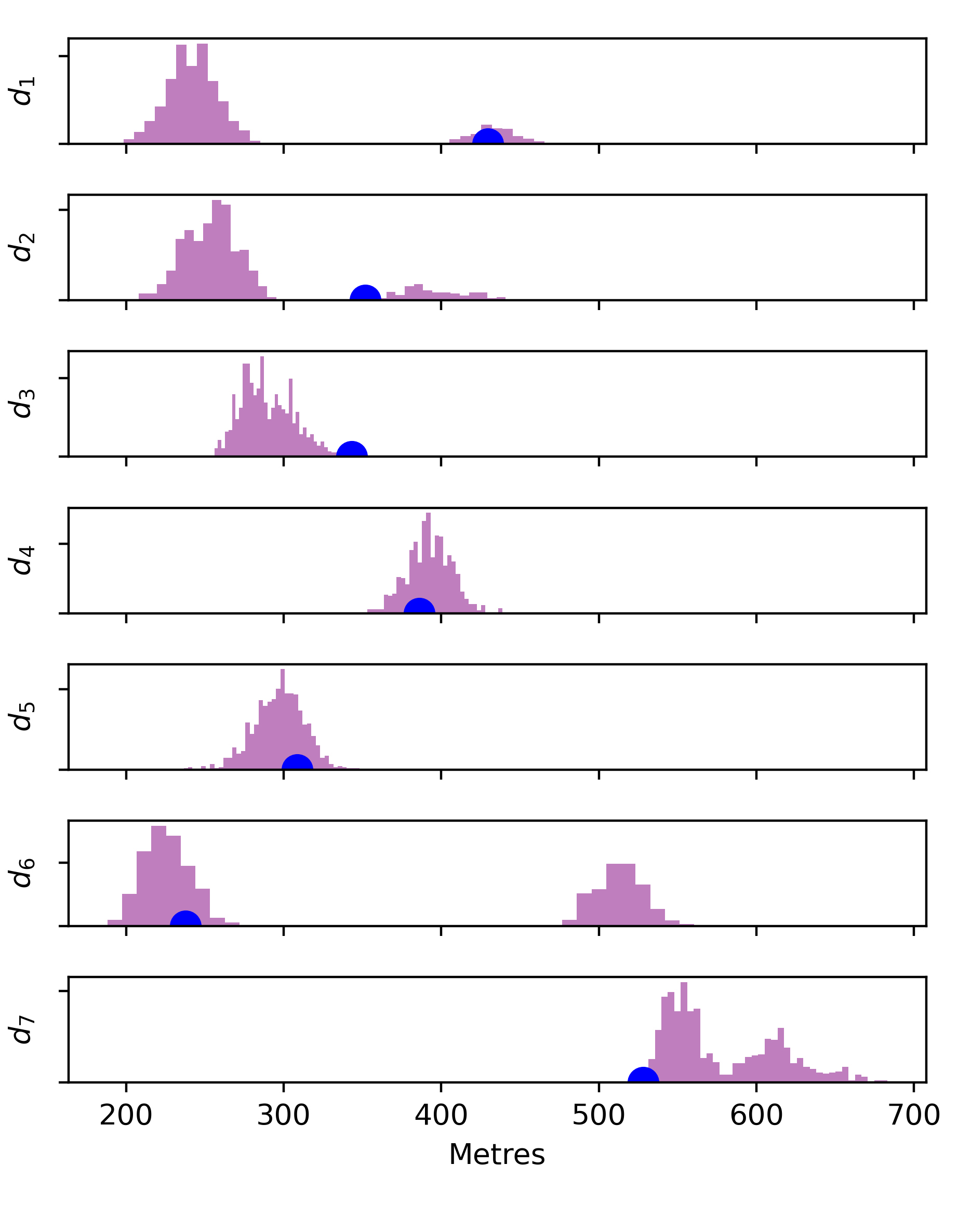}
		\caption{Histograms represent $p(\|x_t - x_{t-1} \|_{e_t}|y_{0:T})$ from FFBSi. Spots represent distances inferred using Viterbi map-matching.}
		\label{fig:synth_hists}
	\end{subfigure}
	
	\caption{Offline particle smoothing vs Viterbi map-matching for synthetic trip across Cambridge.}
	\label{fig:synth_route}
\end{figure}

We tuned the model hyperparameters using offline gradient expectation-maximisation \cite{kantas2015} (running offline FFBSi for the E-step) over 20 routes from the Porto taxi dataset \cite{taxidata} where observations are 15 seconds apart, resulting in values of $p^0 = 0.14$, $\lambda = 0.07/15$, $\beta = 0.05$ and $\sigma_{\text{GPS}} = 5.2$, all edges are discretised to a resolution of 1m.
\par
The true map-matching posterior is analytically intractable and instead we use the approach of \cite{Roth2012} as our \textit{gold standard} for benchmarking: offline forward filtering-backward simulation with a large $N=1000$, i.e. Alg.~\ref{alg:pf} to generate the filtering marginals and Alg.~\ref{alg:ffbsi} for the smoothing particles.

\subsection{Synthetic Data}
In \figurename~\ref{fig:synth_route} we demonstrate the benefits of uncertainty quantification for offline map-matching by comparing our particle based gold standard (offline FFBSi) which is the approach of \cite{Roth2012} against the popular optimisation approach of \cite{Newson2009} on synthetic observations for a trip between the Cambridge Engineering department and the Fort St George pub.
\par

Although the optimisation based approach finds a plausible route (point estimate) it misses out on others that are equally plausible and thus valuable information is lost, this is particularly prevalent when inferring the distances the vehicle travelled, \figurename~\ref{fig:synth_hists}. There is significant uncertainty in both the edges traversed and the distances travelled that is captured by FFBSi but not by the Viterbi algorithm.
\par

All particles generated by FFBSi represent a plausible route - more direct routes are preferred but not overly so - this provides evidence to suggest the model is well suited to difficult dense urban road networks and that the optimal proposal is efficiently generating high probability particles.

In the problem considered here and in \cite{Newson2009, Roth2012} there is no variability in the position of the vehicle along the width of the road - arguably more suitable for dense, urban networks. A less constrained inference problem in e.g. \cite{Gustafsson2002} would also infer the position along the width of the road. The model can then be formulated with Gaussian noise allowing Kalman filter or Rao-Blackwellisation techniques (that are not applicable to our state-space model) can be used to accelerate inference.

\subsection{Real Data}

We now explore the effects of the algorithmic parameters in the online particle smoothers (sample size, lag parameter and number of rejections) for map-matching a route from the Porto taxi dataset \cite{taxidata}. \par

\begin{figure*}
\hspace*{\fill}
\begin{subfigure}[c]{0.48\textwidth}
	\centering
	\includegraphics[width=\textwidth]{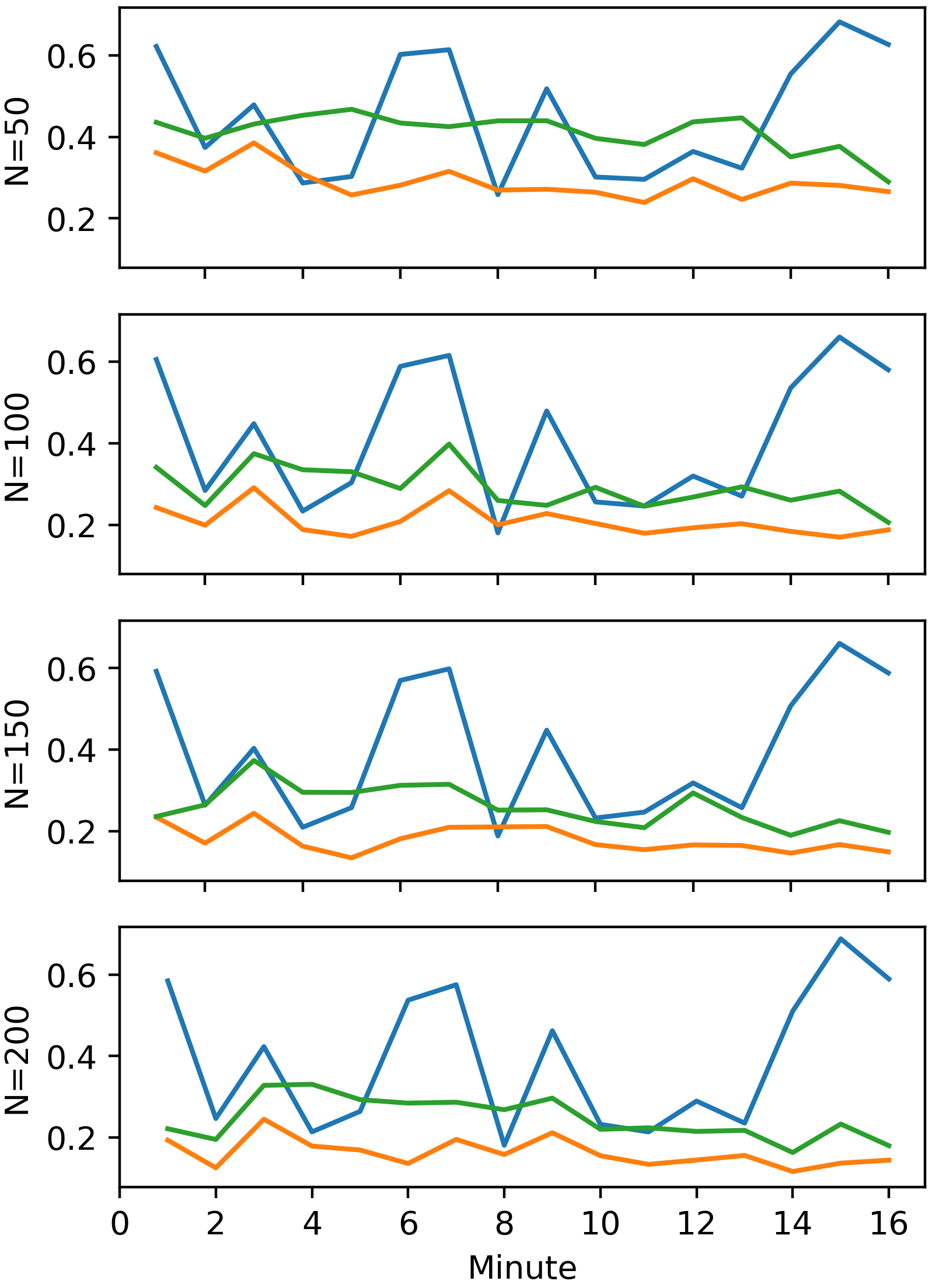}
	\caption{Online Particle Smoother}
	\label{fig:single_route_tv_pf}
\end{subfigure}
\hfill
\begin{subfigure}[c]{0.48\textwidth}
    \centering
    \includegraphics[width=\textwidth]{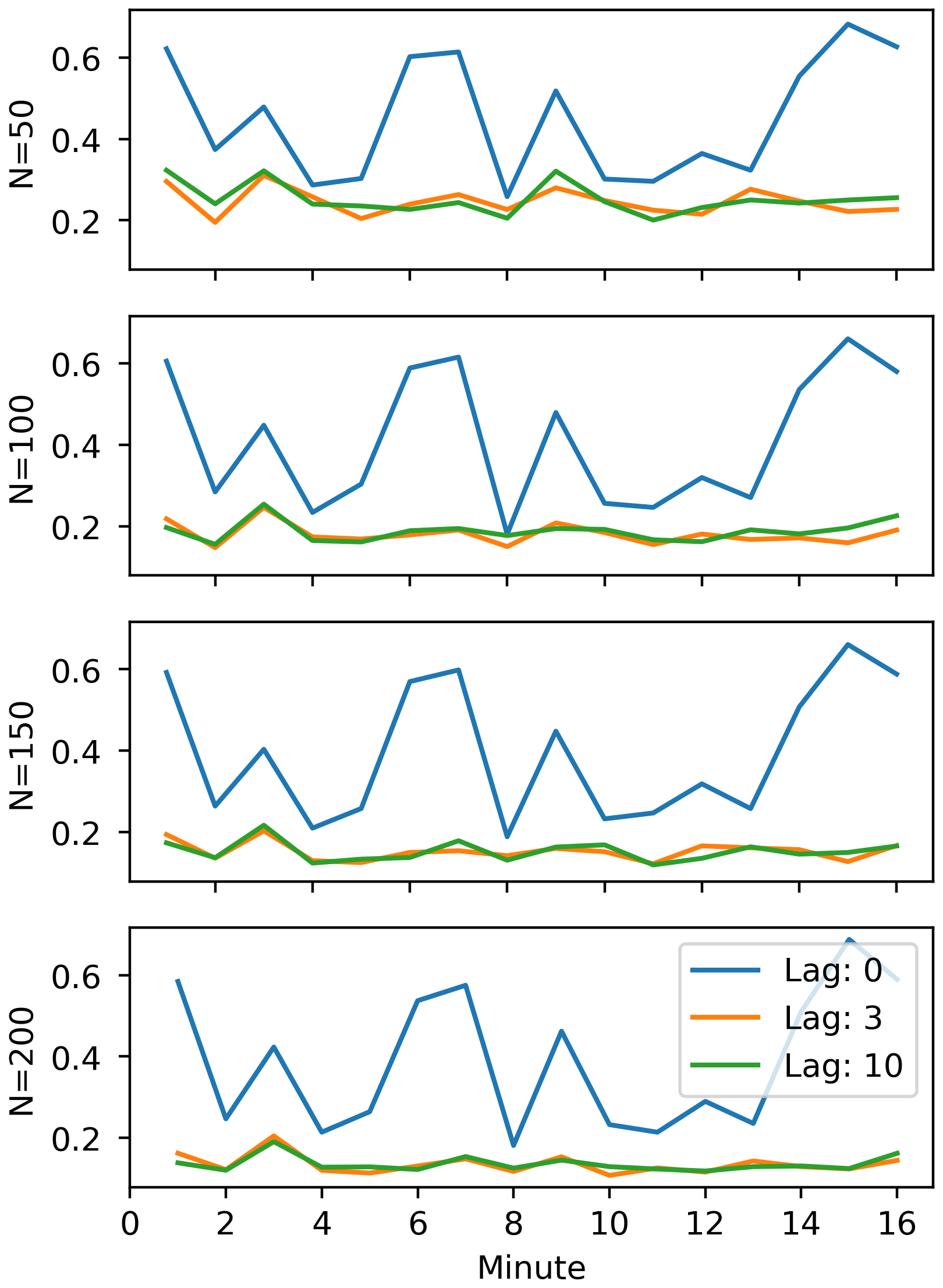}
    \caption{with Backward Simulation.}
    \label{fig:single_route_bsi}
\end{subfigure}
\hspace*{\fill}
\caption{Total variation distance from gold standard (offline FFBSi) for posterior over cumulative distance travelled in each minute of a 16 minute taxi route (observation every 15 seconds) \cite{taxidata}. Simulations averaged over 20 random seeds.}
\label{fig:single_route_comp}
\end{figure*}

In \figurename~\ref{fig:single_route_comp}, we analyse the posterior distribution for the cumulative (road) distance travelled by the taxi each minute. As observations are received every 15 seconds this amounts to expectations averaged over blocks of the full smoothing distribution and thus marginal approximations \eqref{fixed_lag_m} would be insufficient.
For each minute of the trip, we compare particle approximations by calculating the total variation distance over the empirical distributions. This total variation distance is calculated by binning the distance travelled variable, then the empirical distributions are defined over a discrete space and the total variation distance is tractable
\begin{equation*}
    \text{TV}\left(\{d_1^{(i)}\}_{i=1}^{N_1}, \{d_2^{(j)}\}_{j=1}^{N_2} \right)
    = \frac12 \sum_{[a,b] \in \mathcal{D}}
    \left|
    \frac{\sum_{i=1}^{N_1} \mathbb{I}[d_1^{(i)} \in [a,b]]}{N_1}
    -
    \frac{\sum_{j=1}^{N_2} \mathbb{I}[d_2^{(j)} \in [a,b]]}{N_2}
    \right|,
\end{equation*}
where $\mathcal{D}$ represents $[0, \infty)$ discretised into 5 metre width bins and the empirical distances are assumed to be unweighted (although easily adjusted to include weights).

\begin{figure}
    \centering
    \begin{subfigure}{0.49\columnwidth}
    	\captionsetup{width=0.95\textwidth}
        \includegraphics[width=\textwidth]{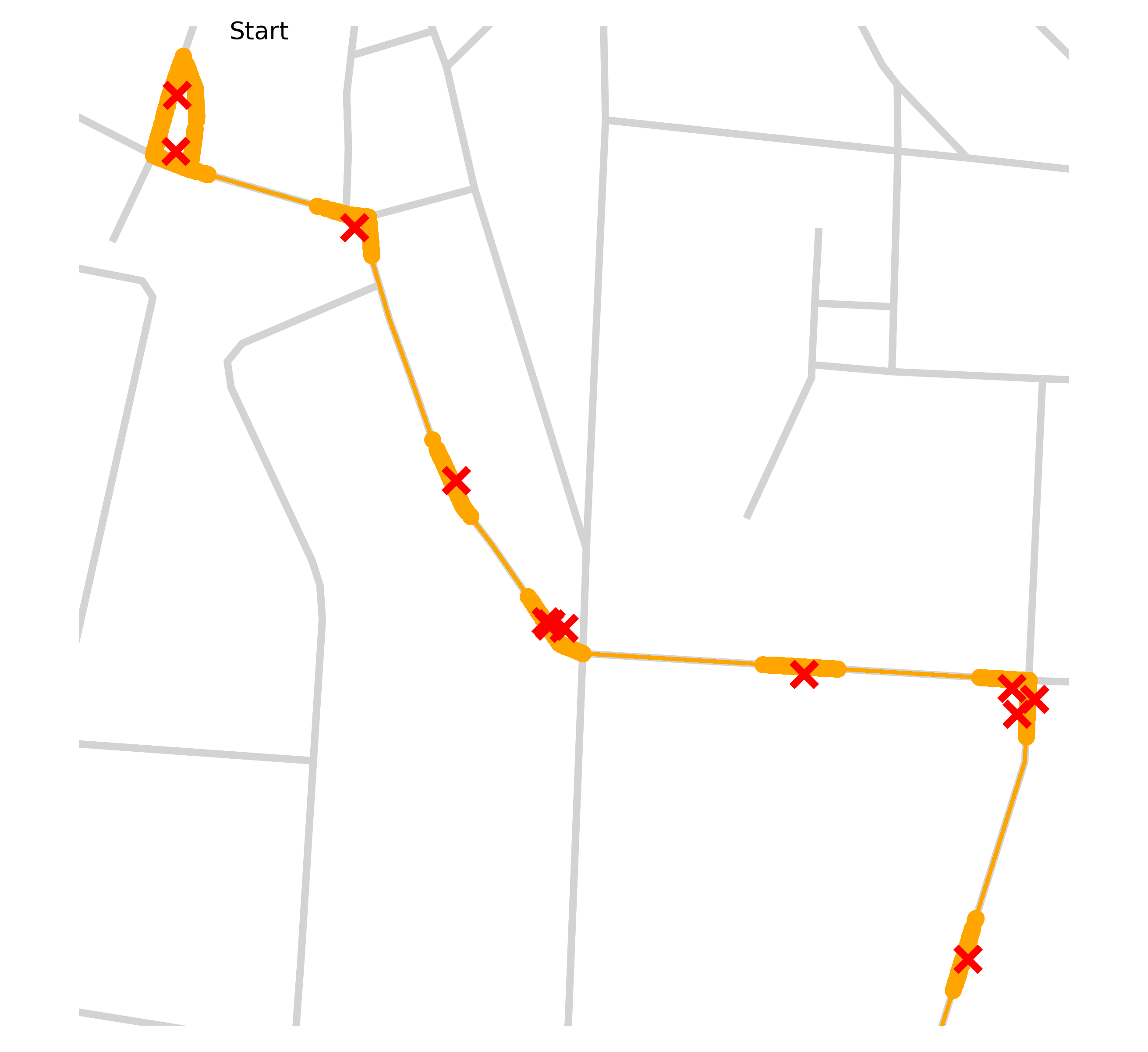}
        \caption{Gold standard, offline FFBSi with $N=1000$.}
        \label{fig:real_ffbsi}
    \end{subfigure}
    \hfill
    \begin{subfigure}{0.49\columnwidth}
    	\captionsetup{width=0.95\textwidth}
        \includegraphics[width=\textwidth]{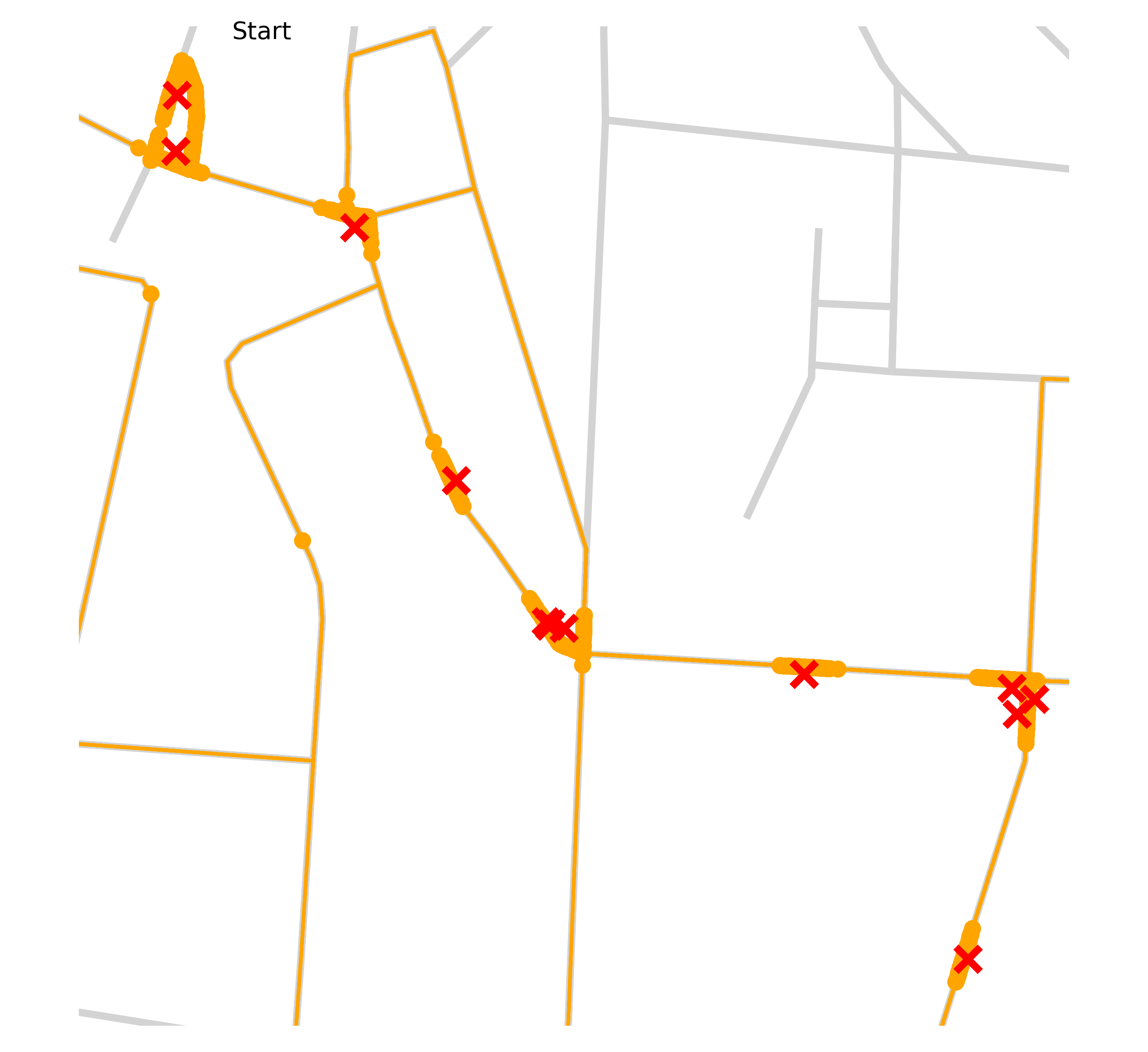}
        \caption{Online particle smoother with $L=0$ and $N=200$. Poor approximation due to small lag.}
        \label{fig:real_fl_lag0}
    \end{subfigure}
    \begin{subfigure}{0.49\columnwidth}
    	\captionsetup{width=0.95\textwidth}
        \includegraphics[width=\textwidth]{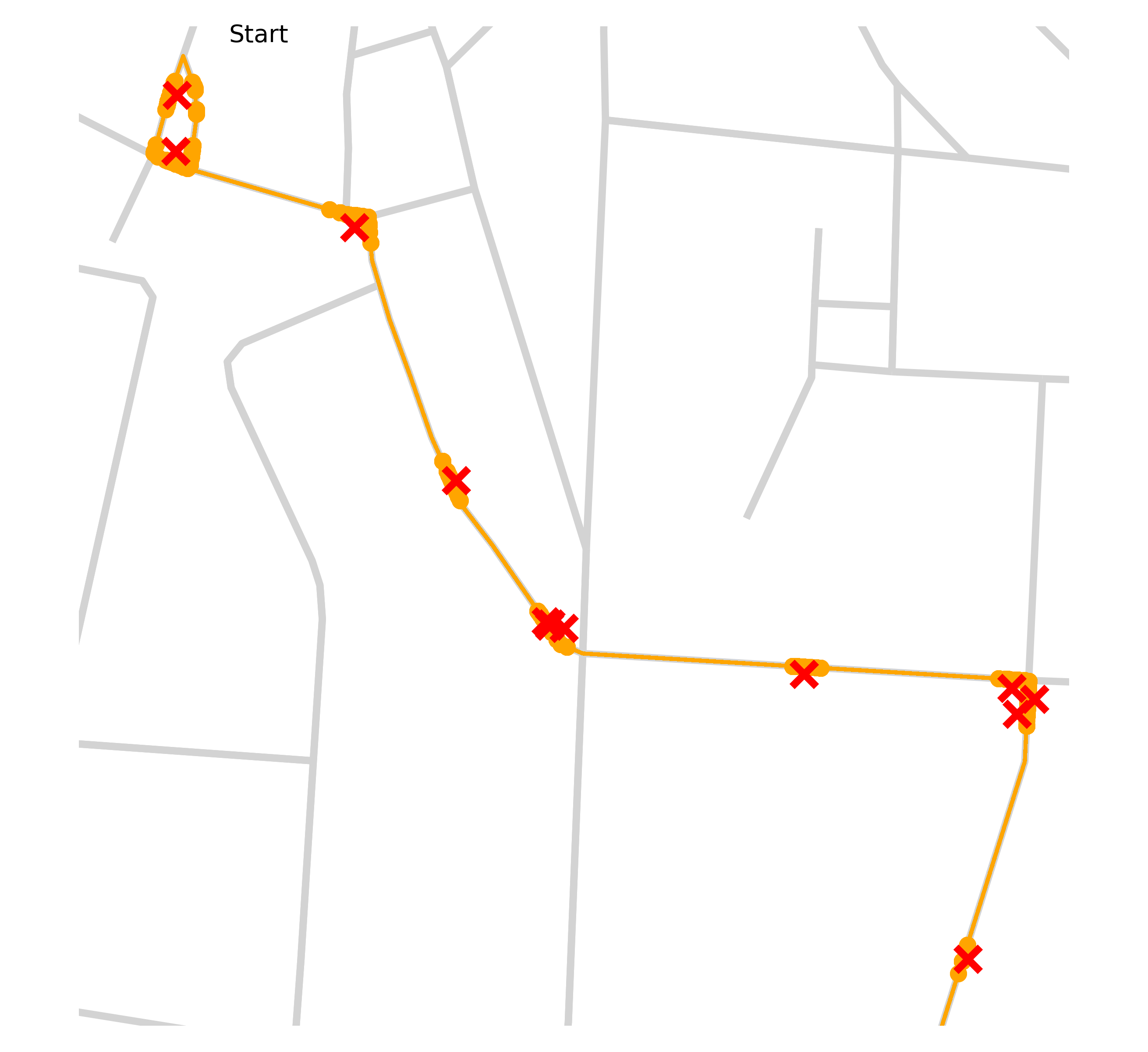}
        \caption{Online particle smoother, $L=10$ and $N=200$. Evidence of path degeneracy due to large lag.}
        \label{fig:real_fl_pf_lag10}
    \end{subfigure}
    \hfill
    \begin{subfigure}{0.49\columnwidth}
    	\captionsetup{width=0.95\textwidth}
        \includegraphics[width=\textwidth]{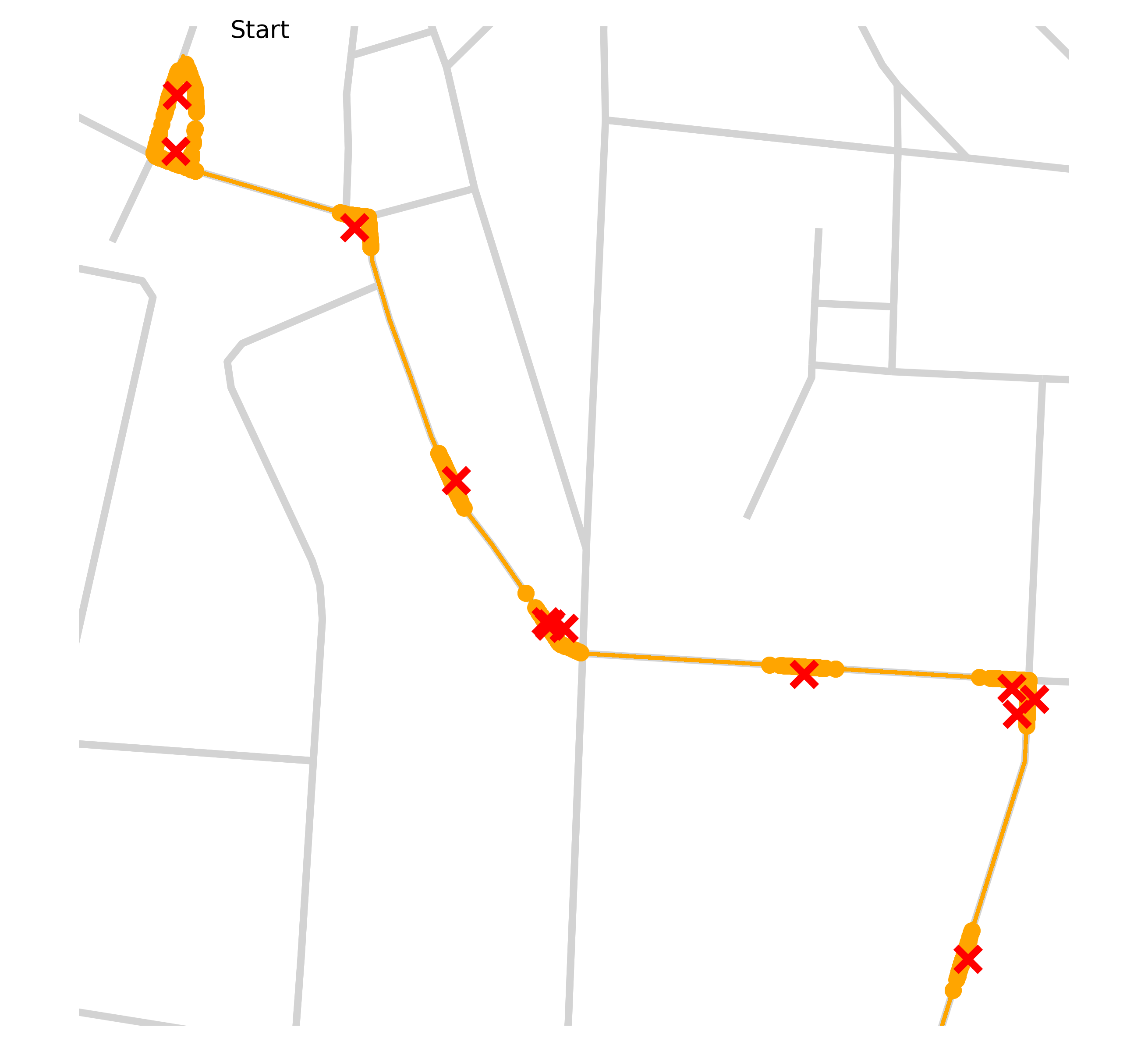}
        \caption{Online particle smoother with backward simulation, $L=10$ and $N=200$.}
        \label{fig:real_fl_bsi_lag10}
    \end{subfigure}
    
    \caption{The start of a route from \cite{taxidata} with various particle approximations.}
    \label{fig:single_start_route}
\end{figure}

\par
\figurename~\ref{fig:single_start_route} depicts the varying algorithmic performance on the start of the route.
\par
We initially observe that setting $L=0$ performs very poorly for all sample sizes. The overly small lag parameter results in a large deviation between the true smoothing distribution \eqref{full_post} and the joint fixed-lag smoothing distribution \eqref{fixed_lag_j}, \figurename~\ref{fig:real_fl_lag0}.
\par
The online particle smoother suffers from path degeneracy for the large lag $L=10$ as observed by a loss of particle diversity (compared to the backward simulation techniques) in \figurename~\ref{fig:real_fl_pf_lag10}. It does however perform well for $L=3$.
\par
The addition of backward simulation completely avoids the issue of path degeneracy, but induces a bias by targeting $p^L(x_{0:T} | y_{0:T})$ rather than $p^L(x_{0:T} | y_{0:T})$, this bias is controllable through the choice of the lag parameter, $L$. We observe that increasing the lag parameter from $L=3$ to $L=10$ does little to improve performance and as such can posit that the distributions \eqref{full_post} and \eqref{fixed_lag_j} are suitably close for $L=3$.
\par
We have not compared numerically to the marginal or additive functional based smoothers described in \sectionautorefname~\ref{sec:rel_work} (such as PaRIS \cite{Olsson2017}) as by definition \eqref{add_func} they are not suitable to expectations over multiple observation times as analysed in \figurename~\ref{fig:single_route_comp} and depicted in \figureautorefname~\ref{fig:arb_vs_stitch}.
\par
A classical particle filter can be recovered by setting $L=\infty$ in the online particle smoother and would suffer path degeneracy to an even greater extent than the online particle smoother with $L=10$.

\begin{figure}
    \centering
    \begin{subfigure}{0.48\columnwidth}
    \centering
        \includegraphics[width=\textwidth]{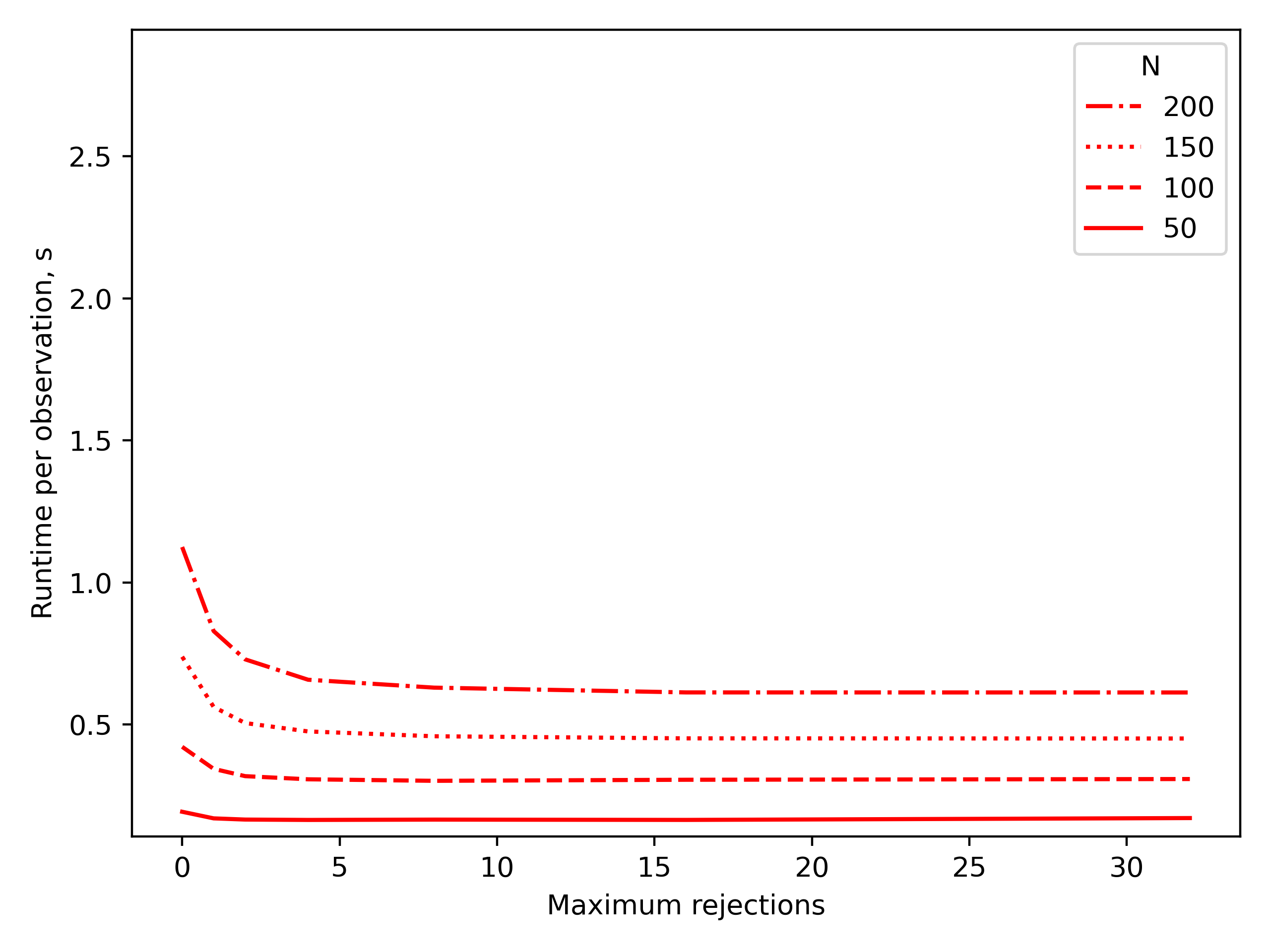}
        \caption{Online Particle Smoother}
        \label{fig:pf_mr}
    \end{subfigure}
    \begin{subfigure}{0.48\columnwidth}
    \centering
        \includegraphics[width=\textwidth]{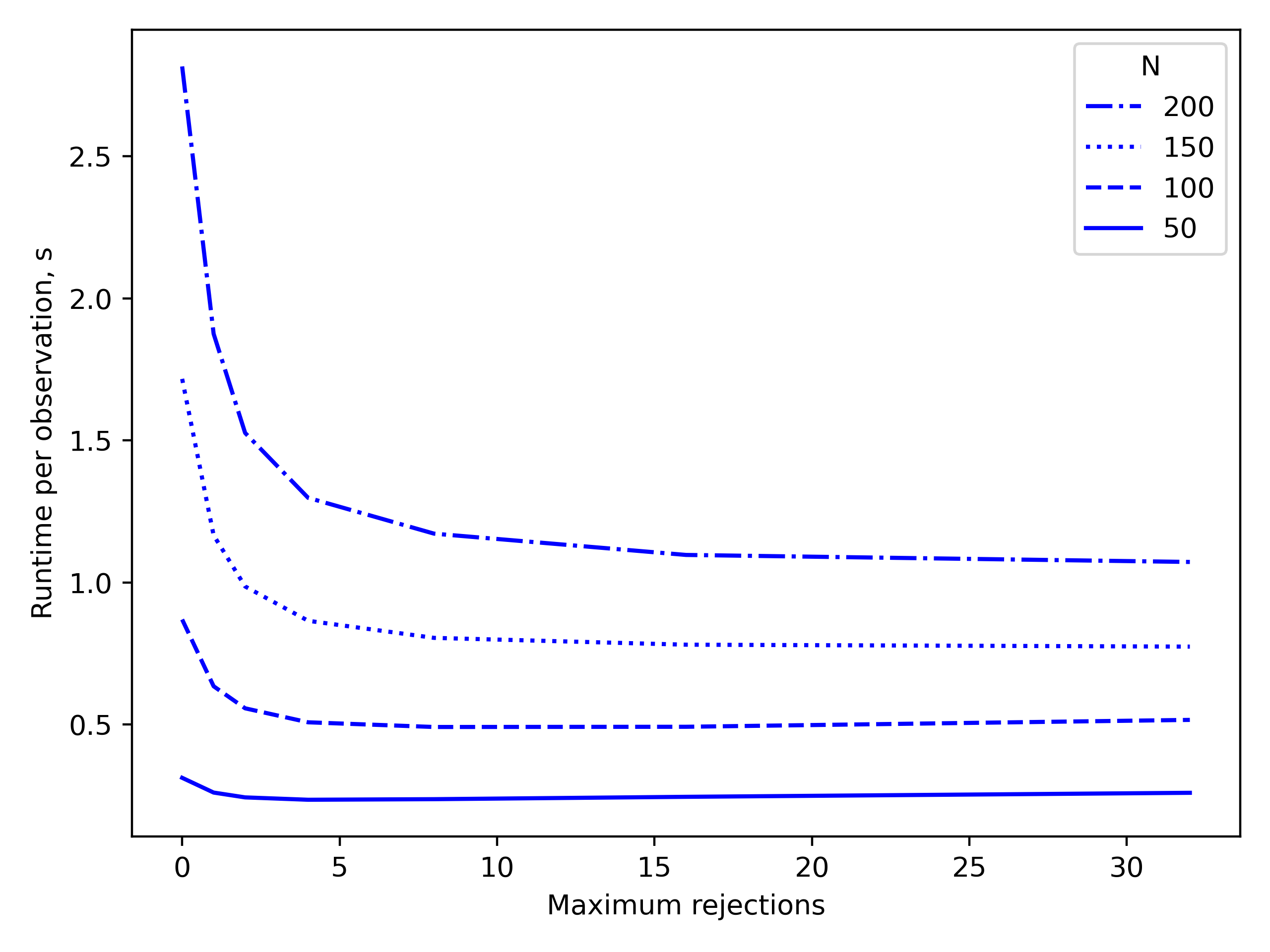}
        \caption{with Backward Simulation.}
        \label{fig:bsi_mr}
    \end{subfigure}
    \caption{Algorithmic runtime vs number of rejections tolerated for $L=3$. Runtimes averaged over 65 observations and 20 random seeds.}
    \label{fig:single_mr_comp}
\end{figure}

Finally, in \figurename~\ref{fig:single_mr_comp} we compare the effect on algorithmic runtime from increasing the maximum number of rejections, $R$, attempted in the hybrid stitching scheme Alg.~\ref{alg:hybrid_stitch}, as well as backward simulation if applicable. Recall that setting $R=0$ recovers the full $O(N^2)$ scheme. For a suitably large number of rejections the runtimes of both algorithms can be seen to increase linearly in $N$.

\section{Discussion}
\label{sec:conc}

In this work, we have developed techniques to efficiently approximate the full joint smoothing distribution or rather the fixed-lag approximation to it in an online setting, this is highly desirable as it permits the online estimation of a range functions that are defined over full trajectories or any subset thereof.  The online particle smoother (Alg. \ref{alg:joint_fl_pf}) comes at the same computational complexity as a classical particle filter, whereas the inclusion of backward simulation (Alg. \ref{alg:joint_fl_ffbsi}) negates the issue of path degeneracy for more difficult models where a large lag parameter is required.
\par

We formulated a state-space model that is specifically designed for urban map-matching and demonstrated the value of particle based uncertainty quantification versus established optimisation based approaches. We then showed that the performance of gold standard offline smoothing FFBSi can be obtained with the online particle smoothers.
\par


The choice of the lag parameter $L$ determines the distance between the distributions $p^L(x_{0:T}|y_{0:T})$ and $p(x_{0:T}|y_{0:T})$, naturally we desire this to be small and therefore $L$ large. The question of how large is a difficult one as it is dependent on the mixing of the state-space model, as discussed in \cite{Olsson2017}.

In practice, the lag parameter can be tuned through preliminary runs on offline or simulated data. This can be done by analysing the sensitivity of smoothing expectations of interest (model-specific) to the value of the lag parameter $L$ and comparing against the true underlying values (in the case of simulated data) or the equivalent expectation from FFBSi. This is investigated for map-matching in \figureautorefname~\ref{fig:single_route_comp}. The tuning of $L$ is perhaps easier for the online particle smoother with backward simulation. The backward simulation completely avoids the issue of path degeneracy and thus error only arises from Monte Carlo variance and the difference between $p^L(x_{0:T} | y_{0:T})$ and $p(x_{0:T} | y_{0:T})$.

An interesting extension would be to investigate the use of a variable lag parameter which is chosen dynamically, as achieved for a function specific version of the marginal fixed-lag particle filter in \cite{Alenlov2019}.
\par

It would be desirable to obtain theoretical results bounding the error induced by the introduced online particle smoothers. We leave this for future work as we anticipate the analysis to be somewhat intricate - combining the work on central limit theorems for particle smoothing such as \cite{DelMoral2010, Douc2011} and the bias induced by a fixed-lag approximation \cite{Olsson2008}. As well as quantifying the impact of using the tractable weights with overlapping coordinate from \ref{subsec:trac_w} as opposed to the optimal but intractable weights in \ref{subsec:intrac_w}.

\par

The realisation of a low-probability transition or observation in the true underlying process can cause degeneracy either at stitching time or in the filtering weights. It is possible to reintroduce particle diversity by applying an MCMC kernel after stitching as in resample-move particle filters \cite{Gilks2001} or particle rejuvenation \cite{Lindsten2015}. The MCMC kernel is applied independently to each particle and must be invariant for the full smoothing distribution $p(x_{0:T}|y_{0:T})$. As we are only concerned with increasing particle diversity rather than taking ergodic averages we need only propose moving a subset of the trajectories, whether that be at stitching time or the latest observation time.


\label{app:ffbsi}
\begin{algorithm}
	\caption{Backward Simulation}
	\begin{algorithmic}[1]\label{alg:ffbsi}
		\item Input weighted marginal filtering samples $\{ \tilde{x}_{t}^{(k)}, \tilde{w}_t^{(k)}\}_{k=1}^N$ for $t=0,\dots,T$, bound $\rho \geq p(x_{t}| x_{t-1})$ and the maximum number of rejections to attempt $R$.
		\STATE Resample
		\begin{align*}
			\left\{ \tilde{x}^{(k)}_{T}, \tilde{w}^{(k)}_{T} \right\}_{k=1}^N
			\rightarrow
			\left\{ x^{(i)}_{T}\right\}_{i=1}^N
		\end{align*}
		\FOR{$t=T-1,\dots,0$}
		\FOR{$i=1,\dots,N$}
		\FOR{$r=1,\dots,R$}
		\STATE Sample $c^* \sim \text{Categorical} \left( \left\{ \tilde{w}_t^{(k)} \right\}_{k=1}^N \right)$
		\STATE Sample $u \sim U(0,1)$
		\IF{$u < p(x_{t+1}^{(i)}|\tilde{x}_{t}^{(c^*)})/\rho$}
		\STATE Accept $c^*$
		\STATE \textbf{break}
		\ENDIF
		\ENDFOR
		\IF{a sample $c^*$ was accepted}
		\STATE Set $x_{t}^{(i)} = \tilde{x}_{t}^{(c^*)}$
		\ELSE
		\STATE Calculate interacting weights and normalise in $k$
		\begin{align*}
			w_t^{(k\leftarrow i)} &\propto
			p(x_{t+1}^{(i)}|\tilde{x}_{t}^{(k)})
			\tilde{w}_t^{(k)}
			\tag*{$k=1,\dots,N$.}
		\end{align*}
		\STATE Sample $c_i \sim \text{Categorical} \left( \left\{ w_t^{(k\leftarrow i)}  \right\}_{k=1}^N \right)$ and set $x_{t}^{(i)} = \tilde{x}_{t}^{(c_i)}$ 
		\ENDIF
		\ENDFOR
		\ENDFOR
		\STATE Output unweighted sample $\left\{x_{0:T}^{(i)}\right\}_{i=1}^N$ approximating $p(x_{0:T}|y_{0:T})$.
	\end{algorithmic}
\end{algorithm}


%
%




\bibliographystyle{IEEEtran}

\bibliography{bibtex/bib/references}

\end{document}